\documentclass{article}

\usepackage{natbib}
\usepackage{float}
\usepackage{arxiv}
\usepackage[utf8]{inputenc} 
\usepackage[T1]{fontenc}    
\usepackage{hyperref}       
\usepackage{booktabs}       
\usepackage{amsfonts}       
\usepackage{amsmath, amssymb}
\usepackage{nicefrac}       
\usepackage{microtype}      
\usepackage{graphicx}
\usepackage{cprotect}
\usepackage{fancyvrb}
\usepackage{listings}
\usepackage{makecell}
\usepackage{ragged2e}
\usepackage{pifont}
\usepackage{array}
\usepackage{multirow}
\usepackage{subcaption}
\usepackage{doi}
\usepackage{tikz}
\usetikzlibrary{arrows.meta, positioning, fit}
\newcolumntype{L}[1]{
  >{\raggedright\arraybackslash
    \hyphenpenalty=10000
    \exhyphenpenalty=10000
  }p{#1}%
}
\newcolumntype{C}[1]{
  >{\centering\arraybackslash
    \hyphenpenalty=10000
    \exhyphenpenalty=10000
  }p{#1}%
}
\newcolumntype{S}[1]{
  >{\raggedright\arraybackslash
    \small\hyphenpenalty=10000
    \exhyphenpenalty=10000
  }p{#1}%
}
\newcolumntype{T}[1]{
  >{\centering\arraybackslash
    \small\hyphenpenalty=10000
    \exhyphenpenalty=10000
  }p{#1}%
}
\newcolumntype{V}[1]{
  >{\ttfamily\scriptsize
    \raggedright\arraybackslash}p{#1}
}
\title{Context operations to architecture modelling output from large language models and evaluation criteria for their use in systems engineering design}

\author{\href{https://orcid.org/0000-0002-8308-3897}{\includegraphics[scale=0.06]{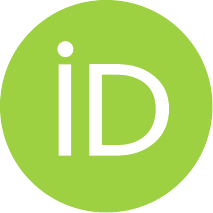}\hspace{1mm}Vinicius Kaster Marini}\thanks{Adjunct professor at the Department of Mechanical Engineering, Centre of Technology, Federal University of Santa Maria, Brazil. You are welcome to contact me through the email above, or you can look for my profiles in LinkedIn and ResearchGate.} \\
  Department of Mechanical Engineering\\
  Centre of Technology \\
  Federal University of Santa Maria\\
  Santa Maria/RS, BR 97105-900, Brazil \\
  \texttt{vinicius.marini@ufsm.br} \\
	\And
	\href{https://orcid.org/0000-0002-2315-0680}{\includegraphics[scale=0.06]{orcid.pdf}\hspace{1mm}Petter Krus} \\
      Section for Fluid Power and Mechatronics\\
      Department of Management and Engineering\\
      Linköping University \\
      SE 581 83, Linköping, Sweden \\
      \texttt{petter.krus@liu.se} \\
}

\renewcommand{\shorttitle}{Context operations with LLMs and evaluation criteria for systems engineering}

\hypersetup{
pdftitle={Context operations with large language models in systems engineering design},
pdfsubject={eess.SY, cs.AI, cs.SE},
pdfauthor={Vinicius Kaster Marini, Petter Krus},
pdfkeywords={large language models, systems engineering, context operations, prompt engineering, generative AI},
}

\begin{document}
\maketitle

\begin{abstract}
	The development of generative artificial intelligence resources enables opportunities of speeding up systems and engineering design work. This contribution introduces a framework of formal operations for assembling context in LLM-based engineering design. This framework involves the assembly of modular context units, including policy prompts, reference units with persistence, and user questions with prompt vectoring. This approach enables the systematic structuring of interactions with generative models. A formal method for evaluating modelling-as-code LLM outputs is also presented, which enables the evaluation of compliance to intent from LLM answers and thereby asses the support from LLMs for systems architecture modelling.
\end{abstract}

\keywords{Large language models \and Systems engineering \and Context operations \and Prompt engineering \and Generative AI}

\section{Introduction} \label{sec:intro}

The complexity of automation in mobility systems draws to human-machine interfaces in operations as a matter of concern for systems engineering (\cite{cummings_AutomationBias_2004}). The implementation of artificial intelligence (AI) to augment control in novel vehicles has implications that extend back to the design process. The realization of gains from AI-enabled and connected smart devices entails added complexity and intricacy in the development of system designs as multi-domain system stacks (\cite{torngren_Complexity_2018,grogan_perception_2021}).

This contribution aims to demonstrate that human-machine collaboration through the system design process can be designed in a way that leverages generative artificial intelligence (GenAI) capabilities. Among many approaches of AI, tools with these capabilities make use of large language models (LLMs) (\cite{brown_language_2020, vaswani_attention_2017}), which are pre-trained over a very large corpus of data over the internet to yield conversational abilities in answering questions. These tools have attracted the attention of system and design engineers (\cite{Krus_LLMSAerospaceICAS_2024, johnsetal_LLMstoMBSE_2024}) for the potential support to early design, a context where other design automation techniques -- including those based on machine learning as well as LLMs -- fell short.

However, their use in engineering and design is subject to challenges regarding their probabilistic approach to content (\cite{teubner_welcome_2023}). While synthesis of assurance arguments makes a potential use case for LLMs upon the amount of paperwork involved, a research report by NASA analyses early explorations with LLMs and highlights their lack of matter-of-factness (\cite{graydon_UsesofLLMs_2025}). Hence, the use of LLM-based tools in engineering and design requires careful review of the LLM outcomes towards design work products (\cite{gomez_LLMs_2024}).  

Then, \textit{how to improve the accuracy of GenAI to design intent towards system architecture modelling}? 

This contribution introduces the theme of LLMs in systems design and engineering within the use case of system architecture modelling, with developing from awareness to the state-of-the-art in LLMs (\cite{marinietal_Human-machine_2025}) towards the systematization of context input for that purpose. Related work involves the following use cases: assistance to systems modelling; assistance in system tools; example\&rule assistance; complementary knowledge; assisted safety-driven methods; assisted reliability\&safety methods; exploration of dependencies; and HMI for design assistance.

\textbf{**Assistance to systems modelling**}: Explorations on using LLMs with systems models demonstrate the early use of developing language models along system modelling frameworks. \cite{camara_Assessment_2023} experiment with prompting at the language models with focus on a single system modelling task, supported by model templates intended to provide exemplars. The integration of modelling frameworks through exemplars helps at extracting useful information to proceed a with a significant part of model-building.

\textbf{**Assistance in system tools**}: LLMs with chatbox tools can work within model-based systems engineering (MBSE) environments such as reported by \cite{dehart_LLM&SysML_2024} and \cite{ johnsetal_LLMstoMBSE_2024}, where the use of LLMs benefits from capturing the modelling framework in the MBSE environment and thereby enables the generation of system models from concept to architecture. This is also the focus in \cite{timperley_assessment_2025}, who enable prompting at coding frameworks interface between the LLM tool and the modelling environment. Here, they add a design element ontology which enables their solution to provide significant support over the synthesis of design specifications. 

\textbf{**Example\&rule assistance**}: \cite{Krus_LLMSAerospaceICAS_2024} explores the use of LLMs within aircraft concept design with support of structured templates and domain-specific rules to generate system configurations onto prompting the LLM to compose and generate system models with considering these inputs. This approach evolves from early contributions by using a preliminary domain question refined to a systematic prompt with topic structure, aiming to convey design intent and composition rules on objects and their mutual relations in the intended model.

\textbf{**Complementary knowledge**}: \cite{baluetal_LLM-RAG_2025} prompt LLMs to generate safety requirements; as they recognize the limitations by LLMs within their own pre-training, they elicit the aid of databases for the language model.  They sample LLM responses at safety-focused prompts and figure the performance of agent-based RAG to generate better accurate responses against design intent. This is also the case with \cite{hanke_AIAugmentedSE_2025}, who make use of RAG database support towards parsing unstructured content in design repositories towards structured content that can be leveraged onto system models.

\textbf{**Safety-driven methods**}: The potential of LLMs on the generative synthesis of design information has not gone unnoticed by the safety and reliability community. \cite{nouri_LLM-Req_2024} make use of LLMs to generate safety requirements for automotive applications, with proposing a pipeline of prompts designed to automate a hazard analysis and risk assessment (HARA) procedure. Another approach is proposed by \cite{elHassani_integratingLLM-FMEA_2024,elHassani_integratingLLM-FMEA_2025}, who focus the processing of relationships in failure modes and effects analysis (FMEA) process with support from product-related data, through crafted prompts, RAG and model fine-tuning.

\textbf{**Assistance to reliability**} \cite{qi_STPA-GPT_2025} make use of LLMs in an elaborate approach to performing systems-theoretical process analysis (STPA) by experimenting with prompt compositions and communication patterns between engineers and LLMs with different degrees of automation including stepwise review. The use of meta-structures to be supported by LLMs, is a characteristic in the application by \cite{chen_trusta_2025}, where trustworthiness derivation trees (TDT) convey hierarchical dependencies between safety claims -- generated by LLMs and curated of purpose-designed user interface -- help the synthesis of assurance cases. 

\textbf{**Exploration of dependencies**}: Other approach for using LLMs in systems design and engineering is the exploration of dependencies within process models. \cite{lipizzi_text_2025} looks to capture dependencies between information concepts through the synthesis of triplets and their vectoring to explore the design space and synthesize it into sentence-based network graphs. Another way to look at information dependencies examines the use of LLMs to generating design structure matrices (DSMs) representing the design space through connections between design objects (\cite{koh_retrieving_2025}).

\textbf{**HMI for design assistance**}: Counter to the perception that the designer is to be automated out of the process, \cite{marinietal_Human-machine_2025} used concept maps to be parsed/splitted to enable sequenced prompt chains on mission design information for aircraft design. \cite{krus_augmenting_2025} explores ways of working with LLMs in different modes of operation: firstly, the direct generation of models from prompting; then, aircraft architecture models generated  through on-the-fly generated application code that instantiates LLM-generated modules; then, through the embedding of LLM API calls in systems design applications.

These use case propositions demonstrate the diversity of situations where LLM-supported GenAI tools could help systems design. At the same time, this contribution aims to further develop knowledge and practice on their use by providing an a formal structured approach to LLM operation, and an example on how it works.

\section{Overview of LLM-based engineering approaches} \label{sec:overview}

The use cases presented in the introduction establish the field of operation towards positioning this contribution is positioned to support systems engineering and design. Figure 1 shows a morphological matrix characterizing the use cases (rows) and the workflow mechanisms (columns) of the approaches identified in the literature. 

\begin{figure}[htbp]
    \centering
    \includegraphics[width=0.9\columnwidth]{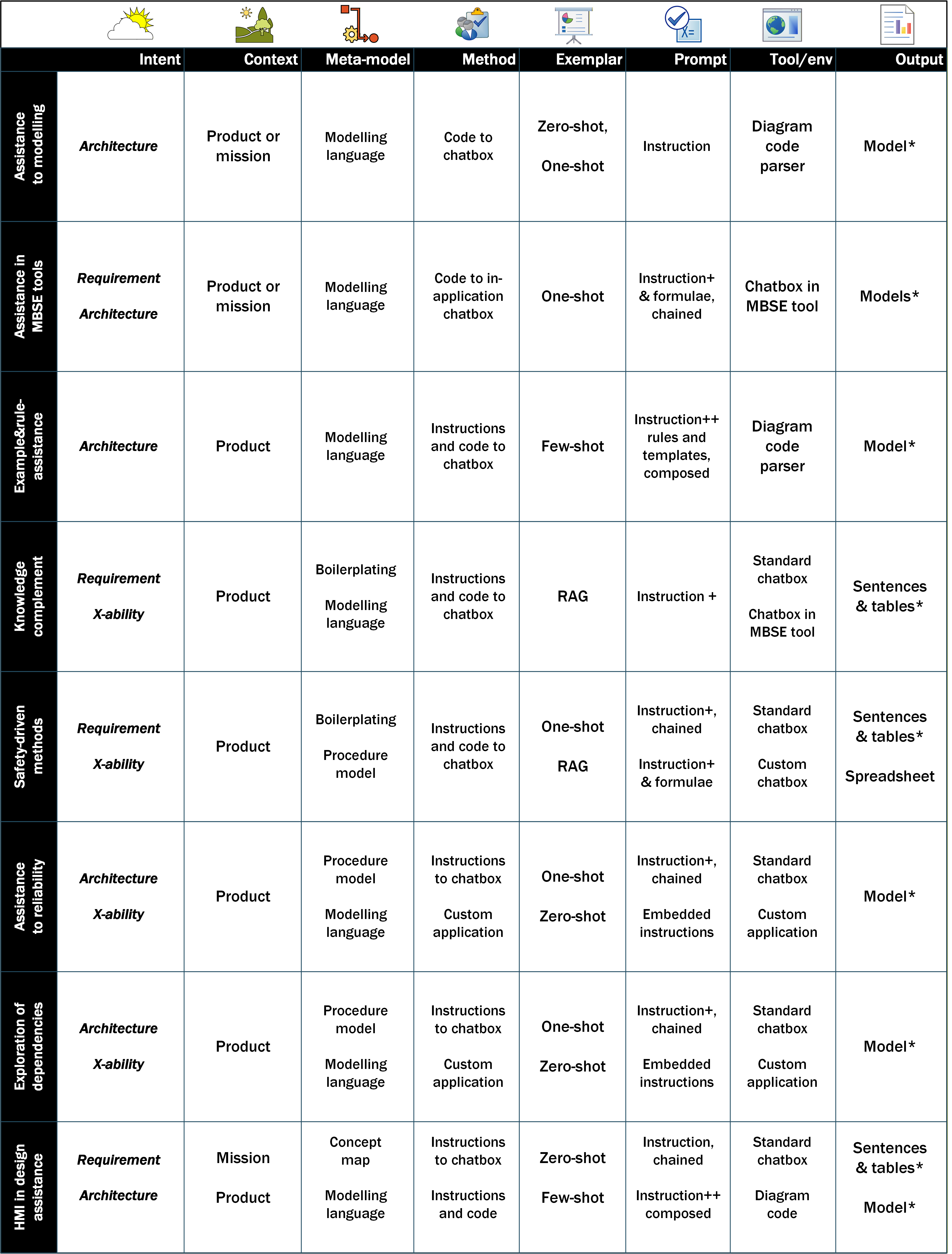}
    \caption{Field of operation and positioning of this contribution}
    \label{fig:field}
\end{figure}

The field overview from Figure 1 displays use cases that demonstrate the performance of  LLMs in delivering outcomes within intended semantic and grammar approaches, often alongside sentence-based outputs. Here, we can see different characteristics of the use cases.

\subsection{Use case design} \label{sec:use-cases}

\textbf{**Intent**}: The intent of using the LLM in each use case regards the proposition of the outcome towards the engineering design process. Design \textit{*requirements*} state properties that the technical system under development shall meet or comply with over its lifecycle. System \textit{*architecture*} assembles system characteristics/elements and their relations onto models conveying properties of the system. \textit{*X-ability*} definitions regard the systematic processing of requirements and architecture to identify actual performance attributes of the technical system.

\textbf{**Context**}: The activity intent involves context in which it addresses product characteristics and missions within various applications.  The use of LLMs in engineering design can address \textit{*Product*} development, when the focus of the approach is to develop a physical system that will be produced for use as part of a given operating context; or, it can address \textit{*Mission*} design, when the focus is to develop the actual operation with physical systems being used within it, which means any physical system will be custom-produced for use within its context.

\textbf{**Meta-model**}: The semantic meta-structure expressing the context information is seen to provide LLMs a route to achieving the intent towards certain engineering context. The formalization of design content for delegating knowledge processing to LLMs works through \textit{*Modelling language*} related to the grammar and \textit{*Procedure model*} related to the design representation. Then, \textit{*Boilerplating*} uses clause/sentence structures to help automate requirement semantics and \textit{*Concept maps*} provide visual representation of how elements and relationships are arranged.

\subsection{Context workloading}

\textbf{**Method**}: Because LLMs are language tools based on large-scale datasets, the approach to requesting the knowledge processing task bears significant influence on its output. The primary method in use by LLM approaches is by standard \textit{*Instructions to chatbox*}, leveraging various prompting techniques in single questions or in conversations comprising a sequence of questions. There is the possibility of prompting with \textit{*Code to chatbox*} as main conveyor of context in association to short requests. Then, the use of \textit{*Custom application*} involves a purpose-specific application where LLM prompts are embedded within its workings and actions are performed by API calls.

\textbf{**Exemplar**}: The use of exemplars is widely recognized as supportive to provide LLMs a better context in regards to the objectives of compliance with processing intent. Users can leverage the methods with exemplars. While \textit{*Zero-shot*} involves no exemplars besides the core question, \textit{*One-shot*} and \textit{*Few-shot*} can involve one or more exemplars, respectively, which convey relevant grammar, semantics, and situations that are relevant to the query. The use of \textit{*RAG*} also helps guiding LLM responses to make answers that are closer to query intent, by having the queried LLM to draw on a domain-specific vector database.

\textbf{**Prompt**}: This regards how the demonstrated use cases frame user input to LLMs. \textit{*Instruction*}-based prompts from the query provide general guidance to how the LLM shall process language to yield its output. Then, \textit{*Formulae*} and \textit{*Rules*} offer more structured control - through the relations between elements are still probabilistic - to get the LLM to yield in compliance to certain occurrence relationship from its training data. There are use cases with \textit{*Templates*}, which show to be useful in providing detailed guidance about expected output formats and characteristics, and \textit{*Chained*} instructions allow for sequenced responses from the buildup of context.

\subsection{Use implementation}

\textbf{**Tool/environment**}: These provide means for the user to interface with requesting information from the LLM, and understanding the outcome of its answer. Most use cases rely on \textit{*Chatbox*} outputs which can be standalone developer-issued tools, or can interoperate with application environments such as \textit{*MBSE tools*}. While standalone or \textit{*Standard*} chatboxes provide incomplete and semi-compliant results, the \textit{*Custom*} integration within applications through API-based routines helps with modelling the input contexts with favourable result to LLM outcomes.

\textbf{**Output**}: The type of output determines the work product delivered by the LLM query workflow. Most use cases engage onto the generation of \textit{*Models*}, which require an underlying grammar the LLM will attend to under its processing; this is possible because developers usually train their LLMs with including language grammars with a large variety of examples. Then, \textit{*Sentences*}, \textit{*Tables*}, and \textit{*Spreadsheets*} involve direct application of natural or coded languages to their processing.

\section{Degraded attention and mitigating factors} \label{sec:attention}

\begin{figure}[htbp]
    \centering
    \includegraphics[width=1\columnwidth]{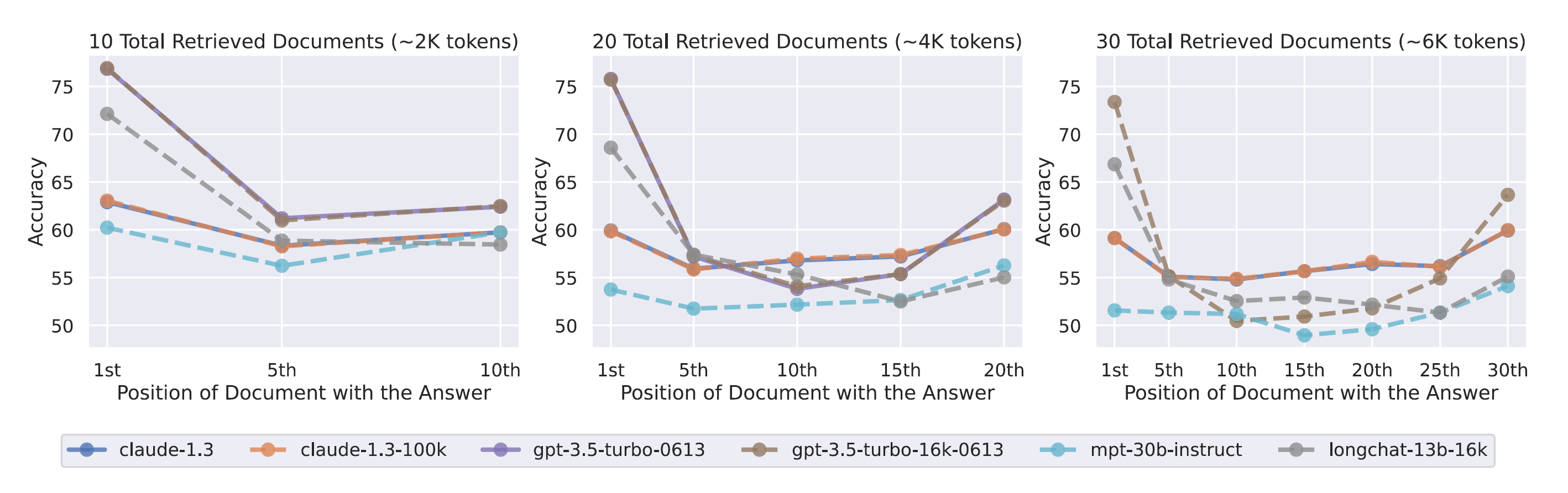}
    \caption{Reliance upon model pre-training for LLM performance.}
    \label{fig:lost-middle}
\end{figure}

Hence, the use of LLMs depends on adding task-related to obtain better work product quality. However, the \textit{transformer} attention mechanism (\cite{vaswani_attention_2017}) cannot equally attend to all tokens in a context workload. This means that forcing the LLM to process multiple sources through long sequences may drive the attention mechanism to lose focus (\cite{liu_lostinmiddle_2017}), thereby reducing the effectiveness in yielding a sufficiently accurate response to the main query. The \textit{lost in the middle} problem is in display by Figure \ref{fig:lost-middle} regarding the context workload length. 

\cite{liu_lostinmiddle_2017} verify the issue becomes more significant in proportion to the length of the assembled context $C_{Q} = \{tk_1..tk_n\}$, considering the actual context capacity of the model - tokens around $tk_1$ and $tk_n$ by both ends of the context workload are more likely to be attended to than those tokens in the middle of the context workload $C_{Q}$. This triggers the need to understand the factors influencing the attention mechanism and how to mitigate them, which motivates use to approach the topic of degraded attention and mitigating factors in the next sections. 

\subsection{Degraded attention factors}

Language processing literature diagnoses a few factors for \textit{transformer}-based models to lose performance about processing the context workload. While there is no reference concerning how these factors play out in systems design and engineering tasks, any technique that uses LLMs as processing resource is liable to these:

\textbf{**Workload length to context capacity**}: the attention mechanism of a specific model has a limit on the number of tokens $\{tk_1..tk_n\}$ in the context workload $C_{Q}$ it can process, a context capacity determined by the pre-training context intake length (\cite{chen2023extendingcontextwindowlarge}). Besides this predetermined overall limit, \cite{workslostinthemiddle_gupte_2025} figure a common characteristic among experimented models, a degradation in recall probability $P[tk_i \in A_{M}]$ as the workload length increases over certain proportion to the total context capacity of a given model.  

\textbf{**Token distance from main query**}: besides the fact that LLMs lose focus on the 'middle' of the context workload $C_{Q}$, \cite{zhang2024middlelanguagemodelsuse} also demonstrate context loss (low $P[tk_i \in A_{M}]$) on tokens $tk_i$ whose position is far from the main query. LLMs run on positional embeddings with higher weights for tokens close to the main query, and lower ones for those farther away from it (\cite{naveed2025comprehensive}). If the main query is located close to the middle, then the attention mechanism will mostly depend on the information by the ends of the workload (\cite{zhang2026positionalfailureslongcontextllms}).

\textbf{**Ambiguity across context units**}: The presence of ambiguous information across different context units along the workload $C_{Q}$ can lead to confusion and misinterpretation by the LLM (\cite{zhang2026positionalfailureslongcontextllms}). This ambiguity, especially when it involves multiple context units around the middle of the workload, can arise from overlapping or conflicting information and further reduce $P[tk_i \in A_{M}]$ because the model has difficulty in determining the desired response. This is also the case when context units are not clearly defined about elements that relate it to the main query.

\textbf{**Evidence complexity**}: \cite{zhang2024middlelanguagemodelsuse} found that the complexity of the evidence at hand within the context workload $C_{Q}$ also plays a role in the degradation of context processing $P[tk_i \in A_{M}]$ towards the answer. Complex context require the so-called 'multi-hop' connection between units across different positions $tk_i$, which is a known challenge for LLMs (\cite{baker2024lostmiddleinbetweenenhancing}). This characteristics means complex evidences require more effort at reasoning about specifics within a main query with respect to the context, which can be challenging for LLMs to maintain and process effectively.

\textbf{**Order-to-structure incoherence**}: The structural incoherence of the context can also lead to degradation in the performance of LLMs. This occurs when the context units are not organized in a coherent manner regarding the work product intent (\cite{li2025ordermattersrethinkingprompt}). This is explained by the behaviour of the attention mechanism, which processes the context workload through sequence-dependent concatenation $C_{Q(i)} = \{tk_1..tk_n\} = ||_{i=1}^{n}\,\, tk_i$. Here, a context workload with units that are misordered on their use to the intended answer can lead to misinterpretation by the attention mechanism, and thereby degrade the response generated by the LLM. 

\subsection{Mitigating degraded attention}
The sources presented in this paper about LLM use cases demonstrate that when context units present information that is complementary and related to the main query. While there are factors inducing attention degradation, others make the effect to  mitigate such mechanism, such as:

\textbf{**Placement about the context ends**}: The attention mechanism of LLMs over the context workload $C_Q = ||_{i=1}^{n}\,\, tk_i$ is subject to the following tendencies about how the LLM reads it (\cite{liu_lostinmiddle_2017}): primacy, for the tokens in the beginning $tk_{i \rightarrow 1}$ of the context workload determine the processing of all subsequent tokens; and recency, for the tokens in its end $tk_{i \rightarrow n}$ point out at the direction to which the model shall attend to. \cite{zhang2024middlelanguagemodelsuse} also point out at the placement of duplicates of relevant context by the ends of the workload to avoid losing that information as reference.

\textbf{**Relevance to the main query**}: The relevance of context units to the main query is a critical factor in mitigating attention degradation. \cite{li2025ordermattersrethinkingprompt} demonstrate that when a higher relationship to the main query from within the pretraining data, atracts attention to specific context. \cite{zhang2024middlelanguagemodelsuse} experiment with placing relevant context units closer to the main query, which they find helps the attention mechanism to focus on those units when looking to improve the quality of the response from the LLM. 

\textbf{**Role-based context positioning**}: this approach involves the strategic placement of context units based on their roles in relation to the main query, with taking advantage of the primacy and recency tendencies upon the sequential processing of the context workload (\cite{liu_lostinmiddle_2017}). \cite{guoetal2024makesgoodorder} demonstrate that the placement of context units based on their roles - policy and rules in $tk_{i \rightarrow 1}$ by the beginning, taking advantage of primacy; then exemplars in $tk_{i \rightarrow n}$ by the end of the workload, taking advantage of recency - is beneficial to the quality of the answer.

\textbf{**Attention-trigger context tagging**}: this approach involves the use of tags or markers to highlight important context units. Attention triggers to context units can involve styling or semantic cues to indicate the importance of a context unit, including its relation to the main query. These can help the attention mechanism to focus on context units about their relationship to the main query, and improve the capability of the LLM to process that particular tagged unit as reference for the answer (\cite{zhang2026positionalfailureslongcontextllms}). 

\textbf{**Reduction of context workload**}: Reducing the number of context units in the workload can help mitigate the effects of attention degradation \cite{workslostinthemiddle_gupte_2025}, because it reduces the competitive demand for the attention capacity by the model. A less diverse context  mitigates degraded attention on both context length and multi-hop requirement effects (\cite{baker2024lostmiddleinbetweenenhancing}) because a less diverse context workload entails less complexity in terms of the number of context units and the relationships between them.

\section{Context assembly levels in LLM use cases} \label{sec:assembly}

A study performed by \cite{graydon_UsesofLLMs_2025} on specific use cases of LLMs in safety engineering finds resulting specifications come incomplete and inaccurate. This led our interest in the factors affecting the quality of the work product $A_M$ delivered by the LLMs with basis in the context workload $C_Q$. Authors such as \cite{camara_Assessment_2023,crabbjonesGenAI_2024} tried the use of LLMs with simple queries to elicit responses from the LLM, and found that examples counting on the query alone yielded insufficient quality regarding the work product delivered by the LLM. 

\cite{gomez_LLMs_2024} and \cite{timperley_assessment_2025}, among others, demonstrate the role of context assembly into building work productas at reasonable quality, which is seen to reduce the amount of rework required to make it compliant to design intent. As section \ref{sec:attention} explores influences to the attention mechanism of LLMs in generic terms, the following sections explore context composition for LLMs in systems design and engineering.

\subsection{Context composition levels}

As section \ref{sec:overview} shows, use cases involve specifics to utilising and assembling supportive context towards the LLM. Besides the overview of working pinciples from Figure \ref{fig:field}, these specifics are outlined along the following characteristics:

\textbf{**Conveyor formats**}: Role, guideline, requirement and directive definitions can involve text prose or specifically-controlled prose when working with safety- and traceability-critical information (\cite{elHassani_integratingLLM-FMEA_2025}); short database records with cross-related information such as requirements, benchmarking and traceable safety information take benefit of csv-, markdown- and other table formats (\cite{conf:geisslerLLMAgent:2024}); and, visual information requires a multi-modal or vision-capable model that can interpret the visual information and relate it to the main query.

\textbf{**Grammar formats**}: These can include formal or informal structures. Natural language syntax is mostly used to convey directives and guidelines, yet can sometimes be used to convey requirements and exemplars (\cite{dehn2025generating}). Formal grammars such as markup (XML, YAML, JSON, etc.) and programming languages (Python, Rust, C++, etc.) in codeblocks (\cite{gomez_LLMs_2024,krus_FluidPowerLLMs_2026}), and programming languages, which convey the relationships between functions and their parameters in a program.

\textbf{**Application adapters**}: These adapters enable the assembly of context and the communication to the LLM and back. Most MBSE use cases leverage application-specific APIs to enable the assembly of context units within the modelling environment and then thits communication to the LLM and back (\cite{dehn2025generating,timperley_assessment_2025}); another way of assembling context workloads is the embedding of  custom code and interfaces in applications through internal routines, which enable LLM calls within their own working environment (\cite{chen_trusta_2025}).

\textbf{**Downstream interpreters**}: In this approach, the forwarding of the context workload to the LLM is carried out by the means of LLM-provider APIs, which get the LLM to generate work products from the exemplars in the same modelling language (\cite{krus_augmenting_2025}). The use of modelling-as-code interpreters can include a single foundation exemplar or a set of exemplars (\cite{Krus_LLMSAerospaceICAS_2024}) that convey the relationships between elements and their properties in the intended work product.

Conveyor and grammar formats enable machine-readable context workload so that the LLM can process it, along application adapters and downstream interpreters that enable LLM outputs to be actionable for a modelling environment. 

\subsection{Context composition strategies}

Some use cases obtain acceptable work products from modelling-as-code context workloads, which enable model-building by downstream interpreters, or by using application adapters within MBSE modelling environments. Both approaches share the following strategies:

\textbf{**Unit composition:**}: The use cases display a diversity of strategies: counting on the pretraining context of the specific LLM (\cite{camara_Assessment_2023}); the cumulation of  downstream question-answer turns at the LLM for in-context learning (\cite{conf:vonheissenGenSysArch:2024}); the use of formal grammar snippets as exemplars supportive to the model-building task (\cite{Krus_LLMSAerospaceICAS_2024}) along with the option to alternate interaction between context workloads for LLM and generated deterministic model-builing code (\cite{krus_FluidPowerLLMs_2026}).

\textbf{**Unit order**}: The placement of references by along the main query (\cite{guoetal2024makesgoodorder}) is driven by the primacy and recency biases. Leveraging the tendencies related to context memory and attention degradation (\cite{liu_lostinmiddle_2017}), an ordered approach to context assembly involves setting rules, precedents, findings and other contextual information following a logical and methodic sequence; examples such as \cite{dehn2025generating} and \cite{krus_augmenting_2025} demonstrate the use of ordered context units - from rules to examples -to improve the quality of the work product delivered by the LLM.

\textbf{**Modelling-as-code**}: This technique involves sequencing the context workload with starting directives that precede $\mu$-template codeblock elements (\cite{krus_augmenting_2025}), and by allowing the modular composition/intake of separate context units (\cite{marinietal_Human-machine_2025}), both techniques making use of models-as-code around the main query. The use of code-based modelling languages is also enabled in examples like those from \cite{timperley_assessment_2025} and \cite{dehn2025generating} that compose the context to LLMs within modelling environments by the means of application adapters.

\textbf{**Context memory**}: The use of context memory involves persistent context units being reused across multiple turns. This can be achieved through the use of RAG databases (\cite{baluetal_LLM-RAG_2025}) or by storing context units in a structured format that can be easily retrieved and reused (\cite{hanke_AIAugmentedSE_2025}). This helps to reduce the amount of context workload required for each query, thereby mitigating attention degradation and improving the quality of the work product delivered by the LLM.

The effectiveness of context composition is determined by how directly each element of the workload conveys information to the LLM, and by how well the context units scaffold the LLM onto processing the work product. 

\section{Model generation parameters} \label{sec:method}

This contribution is part of ongoing development on the use of LLMs as generative AI resource (GenAI) in systems engineering. The understanding of human-machine collaboration with GenAI makes the context, whereas the proposition of knowledge, strategy and applications to leverage the use of GenAI provide directive viewpoint. This study departs from fundamentals in the introduction to this paper and in previous contributions of our own (\cite{Krus_LLMSAerospaceICAS_2024,mariniKrus_PromptComposition_2024}). Then, we aim at a formal approach to context operations in LLM-based systems engineering and design. 

\subsection{Choice of LLM function} \label{sec:choice-llm}

The approach to context operations in this paper first involves formalizing the choice of LLM function and its parametrisation, to configure the LLM emdpoint that will receive the context workload. Then, the context workload assembly is formalized with modedls that present the context units and the operations that enable their composition onto the context workload. 

To understand the context intake functionality of large language models (LLMs), we formalize its workings as a function class $'LLM_X'$, where each specific model plays the role of a function; $'X'$ may be replaced by any particular LLM at the discretion of the user. Here, we use equation clauses to present the functionality and parameters of LLMs, and denote all references to them as $LLM_X$. 

Equation \ref{eq:eq1} approaches LLMs in representing the ability of the professional user to choose one particular model among several available:

\vspace{-10pt}
{\large\begin{equation} \label{eq:eq1}
    \begin{split}
        LLM_X=\vee \space \{&M_{GPT_m}, M_{Claude_m}, M_{Gemini_m}, ..., \\
        &M_{Deepseek_m}, M_{Qwen_m}, M_{Kimi_m}, ..., \\ 
        &M_{Mistral_m}, N_{Command_m}, ...\}
    \end{split}
\end{equation}
}

The first engagement is the choice of operator -- $LLM_X$ function --, which is done with the application by choosing a given individual model operator to answer an intended request. Here, $M$ refers to each being a model, and $m$ refers to one of several versions in a specific model lineup. One can then understand $M$ as an instance of $LLM_X$ which can perform advanced language processing operations, including design and engineering tasks, where $LLM_X$ involves the processing of any chosen $M$ at answering to a request by the user.

\subsection{Parameters of LLM function} \label{sec:param-llm}

Considering the model choice from those in display, each option for $LLM_X$ from equation \ref{eq:eq1} has parameters that can be worked upon. The combination of such parameters in $LLM_X$ can define how the context will be processed. Our contribution considers the following parameters to $LLM_X$:

\begin{itemize}
    \item $M_{lm}:$ \textit{\_model\_size\_} \\ the size of $LLM_X$ on the amonut of parameters in the probability matrix from training,
    \begin{itemize}
    \item $D_{tr}:$ \textit{\_dataset\_size\_} \\ the diversity of sources into the calculation of $M_{lm}$ on which $LLM_X$ is trained,
    \item  $T_{llm}:$ \textit{\_temperature\_} \\ the index to how random the  $LLM_X$ will predict content in its response to a request, and,
    \item  $C_{M(i)_{len}}:$ \textit{\_context\_length\_} \\ the intake capability of $LLM_X$ to receive a context length towards a request,
    \item  $A_{M(i)_{len}}:$ \textit{\_answer\_length\_} \\ the answer length capability of $LLM_X$ to yield content in response to a request,
    \end{itemize}
    \item  $C_{Q(i)_{len}}:$ \textit{\_context\_workload\_} \\ the length of the context assembled towards the call to $LLM_X$.
\end{itemize}

Then, $LLM_X$ can be expressed by Equation \ref{eq:eq2} as a function of the parameters each call relays to the model. The parameters of main interest to our operations are: the \underline{\textit{context length capacity}} $C_{M(i)_{len}}$, the \textit{answer length capacity} $A_{M(i)_{len}}$, and the \textit{assembled context workload} $C_{Q(i)_{len}}$; the first two are properties of $LLM_X$ and the last one is determined from the context operations by the user. 

\vspace{-10pt}
{\large
\begin{equation} \label{eq:eq2}
    \begin{split}
        LLM_X=f(&\overbrace{M_{lm}, \,\, D_{tr}}^{\text{model choice}}, \space \,\, \overbrace{top_p, top_k,T_{llm},...}^{\text{tuning parameters}} \\ 
        &\underbrace{C_{M(i)_{len}}}_{\text{context capacity}}, \,\,\underbrace{A_{M(i)_{len}}}_{\text{answer length}},\underbrace{C_{Q(i)_{len}}}_{\text{context intake}}) \quad \bigg| \quad C_{M(i)_{len}} \geq \{\,C_{Q(i)_{len}} + A_{M(i)_{len}}\,\}
    \end{split}
\end{equation}
}

Each API library has a specific syntax for calling the function and determining its parameters. Specific language models - instances of $LLM_X$ - will also have specific parameter setting requirements for $T_{llm}$, $top_p$ and $top_k$. On our main interest, settings to both $C_{M(i)_{len}}$ and $A_{M(i)_{len}}$ are also model- and provider- specific: within the condition set above, some models set limits at both whereas others allow any proportion between them.

\section{Context operations} \label{sec:operations}

We have engaged onto formalizing the context assembly operations that enable a complete LLM call. Here, we use equation clauses to present context units as to their role onto supporting the LLM's context intake. Our intent is to demonstrate the context-building components towards the enhanced chatbox. Before proceeding with the example, the following context operations are introduced:

\subsection{Generic context formulation} \label{sec:assy}

Context assembly can work through assembling several context units around the main query, whose sequenced combination is intended for ingestion by $LLM_X$ as a single context package. The operator will assemble the context units as available so that $LLM_X$ will perform its internal processing towards the intent. Then, the overall context workload $C_{Q(i)}$ assembly for the '$i$' call will include the following components: 
\begin{itemize}
    \item  $C_{Q(i)}:$ \textit{\_context\_workload\_} \\ the aggregate operated context for an $LLM_X$ function call.
    \begin{itemize}
        \item $C_{up(k,i)} \,\, -\,\, $  \textit{\_upstream\_context\_} in section \ref{sec:policy}: \\ the workload part that is added before the main query to calling $LLM_X$ at each question.
        \item $Q_{P(i)}\,\, -\,\,  $ \textit{\_core\_question\_} from section \ref{sec:core-question} \\ the question statement as elected by the user to call $LLM_X$.
        \item $C_{dn(m,i)} \,\, -\,\, $ \textit{\_downstream\_context\_} in section \ref{sec:references}:  \\  the workload part included after the main query towards $LLM_X$, and,
    \end{itemize}
    \item  $A_{M(i)}:$ \textit{\_answer\_yield\_} \\ the answer yield from $LLM_X$ to yield content in response to a request,
\end{itemize}

The context workload components for $C_{Q(i)}$ will be assembled before - $C_{up(i)}$ - and after -  $C_{dn(i)}$ - the main query, for calling the $LLM_X$ function to the intent expressed in the $Q_{P(i)}$ query. The context workload $C_{Q(i)}$ to the $LLM_X$ function results from assembling context units onto a message; the model internals in $LLM_X$ will process the workload $C_{Q(i)}$ onto an answer $A_{M(i)}$.  Then, equation \ref{eq:eq3a} displays the formulation of each single call to $LLM_X$ with the individual terms to the call. 
\vspace{-12pt}

{\large\begin{equation}\label{eq:eq3a}
    \begin{split}
        \underbrace{C_{up(i)}}_{\text{upstream}} \,\,\,  \Big|\Big| \,\, \underbrace{Q_{P(i)}}_{\text{question}} \,\, \Big|\Big| \underbrace{C_{dn(i)}}_{\text{downstream}} = \quad &C_{Q(i)}\\
        \\
        A_{M(i)}=LLM_X \space \{\,&C_{Q(i)}\,\}
    \end{split}
\end{equation}}

Here, $C_{up(i)}$ is placed first for the upstream context component to set the initial context for the call. Then, the core question unit $Q_{P(i)}$  will express the intent of the context workload. The $C_{dn(i)}$ component will convey imported reference units to provide complementary reference information. Whereas equation \ref{eq:eq3a} only displays context unit grouping per position, there is the opportunity of allowing the intake of several references per context group.

The user will first compose/import the components he expects to forward to $LLM_X$; here, he can add a single component either upstream or downstream of the query, or can add a composition of context units positioned around the main query. Equation \ref{eq:eq3b} represent the context assembly. Here, $'k'$ represents the number of upstream context units and $'m'$ represents the number of downstream context units relative to the main query $Q_{P(i)}$.
\vspace{-12pt}

{\large\begin{equation}\label{eq:eq3b}
    \begin{split}
        C_{up(i)}=  \Big|\Big|^{K}_{k=1} \,\, C_{up(k,i)} \quad &\quad \bigg| \quad \quad C_{dn(i)}=  \Big|\Big|^{M}_{m=1} \,\, C_{dn(m,i)} \quad\, \bigg| \quad\, k, m \in \mathbb{N} \\
        \\
        C_{up(k,i)} \,\, \Big|\Big| \,\, &Q_{P(i)} \,\,\, \Big|\Big| \,\,\, C_{dn(m,i)} = \quad C_{Q(i)}\\
    \end{split}
\end{equation}}

\begin{figure}[htbp]
\centering
    \begin{tikzpicture}[
            lifeline/.style={dashed, thick},
            actbar/.style={draw, fill=gray!20, thick, minimum width=8pt},
            msg/.style={-{Latex[length=6pt]}, thick},
            actor/.style={draw, fill=white, thick, minimum width=40pt, minimum height=20pt, align=center, font=\small},
        ]
        \def\lA{0.5}    
        \def\lB{2.5}    
        \def\lBa{4.5}   
        \def\lC{6.5}    
        \def\lD{8.5}   
        \def\lE{12}
        \def\topY{-0.5}
        \def\botY{-9.0}
        \def\tI{-2}
        \def\tII{-2.75}
        \def\tIII{-3.5}
        \def\tIIIa{-4.25}
        \def\tIIIb{-5}
        \def\tIIIc{-5.75}
        \def\tIV{-6.5}
        \def\tV{-7.25}
        \def\tVI{-8.0}
        \node[actor] at (\lA, \topY) {User};
        \node[actor] at (\lB, \topY) {$C_{up(k,i)}$};
        \node[actor] at (\lBa, \topY) {$Q_{P(i)}$};
        \node[actor] at (\lC, \topY) {$C_{dn(m,i)}$};
        \node[actor] at (\lD, \topY) {$\in C_{Q(i)}$};
        \node[actor] at (\lE, \topY) {$LLM_X$};
        \draw[lifeline] (\lA, \topY-0.5) -- (\lA, {\botY+0.9});
        \draw[lifeline] (\lB, \topY-0.5) -- (\lB, {\botY+0.9});
        \draw[lifeline] (\lBa, \topY-0.5) -- (\lBa, {\botY+0.9});
        \draw[lifeline] (\lC, \topY-0.5) -- (\lC, {\botY+0.9});
        \draw[lifeline] (\lD, \topY-0.5) -- (\lD, {\botY+0.9});
        \draw[lifeline] (\lE, \topY-0.5) -- (\lE, {\botY+0.9});
        \foreach \lbl/\y in {
            {import}/\tI,
            {$||\, C_{up(k,i)} \rightarrow$}/\tII,
            {$> C_{Q(i)} (I) $}/\tIII,
            {compose}/\tIIIa,
            {$||\, Q_{P(i)} \rightarrow$}/\tIIIb,
            {$> C_{Q(i)} (II)$}/\tIIIc,
            {import}/\tIV,
            {$|| \, C_{dn(m,i)} \rightarrow$}/\tV,
            {$> C_{Q(i)} (III)$}/\tVI,
            }{
            \node[font=\footnotesize\itshape, anchor=east] at (0.5, \y) {\lbl};
        }
        \foreach \y in {\tI, \tII, \tIII, \tIIIa, \tIIIb, \tIIIc, \tIV, \tV, \tVI}{
            \draw[gray!25, thin] (0.5, \y) -- (12, \y);
        }
        \draw[->>] (\lA, \tI) -- node[above, font=\scriptsize]{References} (\lB, \tI);
        \draw[msg] (\lB, \tII) -- node[above, fill=white, text opacity=1, font=\scriptsize]{Upstream units $C_{up(k,i)}$} (\lD, \tII);
        \draw[->>] (\lA, \tIIIa) -- node[above, fill=white, text opacity=1, font=\scriptsize]{Query statement} (\lBa, \tIIIa);
        \draw[msg] (\lBa, \tIIIb) -- node[above, fill=white, text opacity=1, font=\scriptsize]{Core question $Q_{P(i)}$} (\lD, \tIIIb);
        \draw[->>] (\lA, \tIV) -- node[above, font=\scriptsize]{References} (\lC, \tIV);
        \draw[msg] (\lC, \tV) -- node[above, fill=white, text opacity=1, font=\scriptsize]{Downstream $C_{dn(m,i)}$} (\lD, \tV);
        \draw[msg] (\lD, \tIII) -- node[left, pos=0, anchor=east, fill=white, text opacity=1, font=\normalsize]{$||_{k=1}^{K} \, C_{up(k,i)}$} (\lE, \tIII);
        \draw[msg] (\lD, \tIII) -- node[above, font=\scriptsize]{eq. (4)} (\lE, \tIII);
        \draw[msg] (\lD, \tIIIc) -- node[left, pos=0, anchor=east, fill=white, text opacity=1, font=\normalsize]{$||\, Q_{P(i)}$} (\lE, \tIIIc);
        \draw[msg] (\lD, \tIIIc) -- node[above, font=\scriptsize]{eq. (4)} (\lE, \tIIIc);
        \draw[msg] (\lD, \tVI) -- node[left, pos=0, anchor=east, fill=white, text opacity=1, font=\normalsize]{$||_{m=1}^{M} \, C_{dn(m,i)}$} (\lE, \tVI);
        \draw[msg] (\lD, \tVI) -- node[above, font=\scriptsize]{eq. (4)} (\lE, \tVI);
        \draw[thick, rounded corners=4pt]
            (-1.6, 0.65) rectangle (13.6, \botY+0.3);
        \node[font=\small\bfseries, anchor=north west] at (-1.6, 0.65)
            {Ancillary context units and query statement: $C_{up(k,i)}, Q_{P(i)}, C_{dn(m,i)} \in C_{Q(i)}$};
    \end{tikzpicture}
    \caption{Generic assembly operations of ancillary context units for a single $LLM_X$ call.}
    \label{fig:context-sum-diagram}
\end{figure}
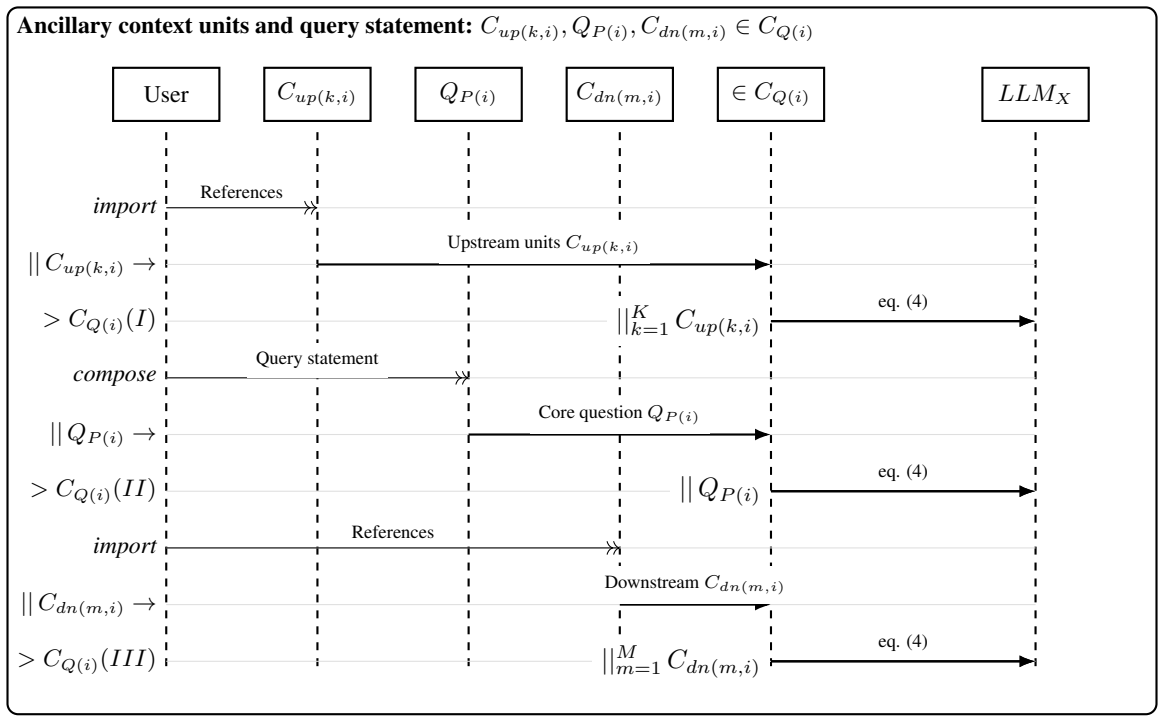

Then, the user can import or compose several context units to be positioned before and after the main query, and the application will assemble them into a single context workload $C_{Q(i)}$ to be sent to $LLM_X$ as in display by Figure \ref{fig:context-sum-diagram}, which displays the context assembly process. The model will process the workload $C_{Q(i)}$ onto an answer $A_{M(i)}$ from its internals.

The assembly of context units from string variables results from the concatenation of individual strings into a single string variable that will carry the full context workload to $LLM_X$. This concatenation of context units is order-dependent upon the alignment between grammar structures in the input and language-processing capabilities by generativce resources such as LLMs. However, the attention mechanism within LLMs has a characteristic of reading the unified context workload from the beginning and from the end of the string, with a tendence to losing attention to the content by the middle of the string. The \textit{lost in the middle} problem as identified by \cite{liu_lostinmiddle_2017} .

\subsection{Role-focused units onto single-call} \label{sec:context-assy}

Role-focused units contain tokenized content attending to distinct purposes in relation to the main query, whose sequenced combination is intended for ingestion by $LLM_X$ as a single context package. The operator will assemble the context units as available so that $LLM_X$ will perform its internal processing towards the intent. Then, the overall context workload $C_{Q(i)}$ assembly for the '$i$' call will include the following components: 
\begin{itemize}
    \item  $C_{Q(i)}:$ \textit{\_context\_workload\_} \\ the aggregate operated context for an $LLM_X$ function call.
    \begin{itemize}
        \item $R_{P(i)} \,\, -\,\, $  \textit{\_policy\_rulework\_} in section \ref{sec:policy}: \\ the resulting rulework prompt with directives to calling $LLM_X$ at each question.
        \item $P_{P(i)} \,\, -\,\, $ \textit{\_context\_reference\_} in section \ref{sec:references}:  \\  a context unit to recur at every call within '$r$' calls of the context memory $C_{mem}$, and,
        \item $Q_{P(i)}\,\, -\,\,  $ \textit{\_core\_question\_} from section \ref{sec:core-question} \\ the question statement as elected by the user to call $LLM_X$.
    \end{itemize}
    \item  $A_{M(i)}:$ \textit{\_answer\_yield\_} \\ the answer yield from $LLM_X$ to yield content in response to a request,
\end{itemize}

The context workload components for $C_{Q(i)}$ shall follow certain order in order to maximise the compliance by the $LLM_X$ function to the intent of the query. The context workload $C_{Q(i)}$ to the $LLM_X$ function results from assembling these context modules onto a message to the model. The model will process the workload $C_{Q(i)}$ onto an answer $A_{M(i)}$ from its internals.  Then, Equation \ref{eq:eq4a} displays the formulation of each single call to $LLM_X$. 
\vspace{-12pt}

{\large\begin{equation}\label{eq:eq4a}
    \begin{split}
        \underbrace{R_{P(i)}}_{\text{policy}} \,\,\, \Big|\Big| \underbrace{P_{P(i)}}_{\text{reference}} \,\, \Big|\Big| \,\, \underbrace{Q_{P(i)}}_{\text{question}} = \quad &C_{Q(i)}\\
        \\
        A_{M(i)}=LLM_X \space \{\,&C_{Q(i)}\,\}
    \end{split}
\end{equation}}

\begin{figure}[htbp]
\centering
    \begin{tikzpicture}[
            lifeline/.style={dashed, thick},
            actbar/.style={draw, fill=gray!20, thick, minimum width=8pt},
            msg/.style={-{Latex[length=6pt]}, thick},
            actor/.style={draw, fill=white, thick, minimum width=40pt, minimum height=20pt, align=center, font=\small},
        ]
        \def\lA{0.5}    
        \def\lB{2.5}    
        \def\lBa{4.5}   
        \def\lC{6.5}    
        \def\lD{8.5}   
        \def\lE{12}
        \def\topY{-0.5}
        \def\botY{-9.0}
        \def\tI{-2}
        \def\tII{-2.75}
        \def\tIII{-3.5}
        \def\tIIIa{-4.25}
        \def\tIIIb{-5}
        \def\tIIIc{-5.75}
        \def\tIV{-6.5}
        \def\tV{-7.25}
        \def\tVI{-8.0}
        \node[actor] at (\lA, \topY) {User};
        \node[actor] at (\lB, \topY) {$R_{P(n,i)}$};
        \node[actor] at (\lBa, \topY) {$P_{P(p,i)}$};
        \node[actor] at (\lC, \topY) {$Q_{P(q,i)}$};
        \node[actor] at (\lD, \topY) {$\in C_{Q(i)}$};
        \node[actor] at (\lE, \topY) {$LLM_X$};
        \draw[lifeline] (\lA, \topY-0.5) -- (\lA, {\botY+0.9});
        \draw[lifeline] (\lB, \topY-0.5) -- (\lB, {\botY+0.9});
        \draw[lifeline] (\lBa, \topY-0.5) -- (\lBa, {\botY+0.9});
        \draw[lifeline] (\lC, \topY-0.5) -- (\lC, {\botY+0.9});
        \draw[lifeline] (\lD, \topY-0.5) -- (\lD, {\botY+0.9});
        \draw[lifeline] (\lE, \topY-0.5) -- (\lE, {\botY+0.9});
        \foreach \lbl/\y in {
            {import}/\tI,
            {$||\, R_{P(n,i)} \rightarrow$}/\tII,
            {$> C_{Q(i)} (I) $}/\tIII,
            {compose}/\tIIIa,
            {$||\, P_{P(p,i)} \rightarrow$}/\tIIIb,
            {$> C_{Q(i)} (II)$}/\tIIIc,
            {import}/\tIV,
            {$|| \, Q_{P(q,i)} \rightarrow$}/\tV,
            {$> C_{Q(i)} (III)$}/\tVI,
            }{
            \node[font=\footnotesize\itshape, anchor=east] at (0.5, \y) {\lbl};
        }
        \foreach \y in {\tI, \tII, \tIII, \tIIIa, \tIIIb, \tIIIc, \tIV, \tV, \tVI}{
            \draw[gray!25, thin] (0.5, \y) -- (12, \y);
        }
        \draw[->>] (\lA, \tI) -- node[above, font=\scriptsize]{References} (\lB, \tI);
        \draw[msg] (\lB, \tII) -- node[above, fill=white, text opacity=1, font=\scriptsize]{Policy units $R_{P(n,i)}$} (\lD, \tII);
        \draw[->>] (\lA, \tIIIa) -- node[above, fill=white, text opacity=1, font=\scriptsize]{Query statement} (\lBa, \tIIIa);
        \draw[msg] (\lBa, \tIIIb) -- node[above, fill=white, text opacity=1, font=\scriptsize]{Reference units $P_{P(p,i)}$} (\lD, \tIIIb);
        \draw[->>] (\lA, \tIV) -- node[above, font=\scriptsize]{References} (\lC, \tIV);
        \draw[msg] (\lC, \tV) -- node[above, fill=white, text opacity=1, font=\scriptsize]{Question units $Q_{P(m,i)}$} (\lD, \tV);
        \draw[msg] (\lD, \tIII) -- node[left, pos=0, anchor=east, fill=white, text opacity=1, font=\normalsize]{$||_{n=1}^{N} \, R_{P(n,i)}$} (\lE, \tIII);
        \draw[msg] (\lD, \tIII) -- node[above, font=\scriptsize]{eq. (5)} (\lE, \tIII);
        \draw[msg] (\lD, \tIIIc) -- node[left, pos=0, anchor=east, fill=white, text opacity=1, font=\normalsize]{$||_{p=1}^{P} \, P_{P(p,i)}$} (\lE, \tIIIc);
        \draw[msg] (\lD, \tIIIc) -- node[above, font=\scriptsize]{eq. (5)} (\lE, \tIIIc);
        \draw[msg] (\lD, \tVI) -- node[left, pos=0, anchor=east, fill=white, text opacity=1, font=\normalsize]{$||_{q=1}^{Q} \, Q_{P(q,i)}$} (\lE, \tVI);
        \draw[msg] (\lD, \tVI) -- node[above, font=\scriptsize]{eq. (5)} (\lE, \tVI);
        \draw[thick, rounded corners=4pt]
            (-1.6, 0.65) rectangle (13.6, \botY+0.3);
        \node[font=\small\bfseries, anchor=north west] at (-1.6, 0.65)
            {Role-based context units and query statement: $C_{Q(i)}$};
    \end{tikzpicture}
    \caption{Role-focused assembly operations to context units for a single $LLM_X$ call.}
    \label{fig:context-role-diagram}
\end{figure}
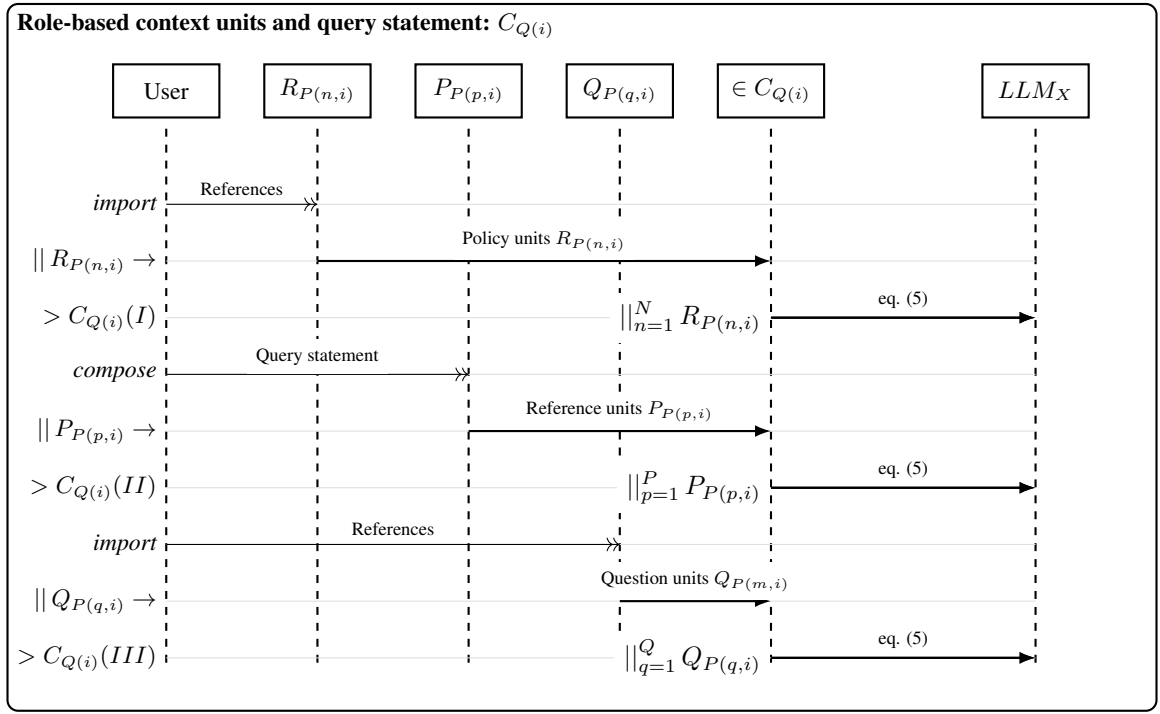

Here, $R_{P(i)}$ is placed first for the policy component to set rules over the whole of the call. The $P_{P(i)}$ component will convey imported reference units to provide complementary guidelines towards the answer. Then, the core question unit $Q_{P(i)}$ can include exemplars such as $\mu$-Templates (\cite{krus_FluidPowerLLMs_2026}) and specifics besides the actual question, and then take advantage of recency (\cite{liu_lostinmiddle_2017}) to maximise the influence of intent-related statements in the query. 

\subsubsection{Rule and policy operations} \label{sec:policy}

One can set policy context units to help steer the workings of $LLM_X$ function to yield an answer to a closer approximation to the query intent. These context units will work as rules of engagement which will drive the $LLM_X$ function to process the core question along the given context of composed and imported units. These context units can be individually composed by the user at anytime regarding its intent towards operating the $LLM_X$ function. Then, the professional user can compose the following string units:
\begin{itemize}
    \item $R_{P(i)}:$  \textit{\_policy\_rulework\_} \\ the rulework prompt with directives and guidelines to calling $LLM_X$.
    \begin{itemize}
        \item $G_{sp}:$ \textit{\_global\_prompt\_} \\ a recurring prompt to every call that provides directives of engagement,
        \item $B_{sp(i)}:$  \textit{\_boundary\_prompt\_} \\ a context memory prompt with complementary guidelines, and,
        \item $R_{P(n,i)}:$  \textit{\_policy\_context\_} \\ additional policy context units for a single call to $LLM_X$.
    \end{itemize}
\end{itemize}

The user will first compose/import the $R_P$ components he expects to forward to $LLM_X$; here, he can add policy components with $n$ policy context units towards setting the policy component $R_{P}$. A context operation regarding policy assembles the $R_{P(i)}$ component with several units $R_{P(n,i)}$, and sends it to $LLM_X$ at the request of the user. The user can compose/import the global prompt $G_{sp}$ to request the LLM to follow a set of rules, without adding a boundary prompt. Here, the call to $LLM_X$ takes place by the (I) message line, and $G_{sp}$ will be forwarded at all times. 
\vspace{-12pt}

{\large\begin{equation}\label{eq:eq5}
	\underbrace{G_{sp}}_{\text{global}} \in R_{P(i)} \quad \quad \bigg| \quad \quad \underbrace{B_{sp(i)}}_{\text{boundary}} \in R_{P(i)}
\end{equation}}

The user can also compose/import the boundary prompt $B_{sp}$ as complementary guidance to $LLM_X$ alongside $G_{sp}$ to the $R_{P(i)}$ policy component. Here, the call to $LLM_X$ takes place by the (II) message line, and $B_{sp}$ context component can be forwarded along the $G_{sp}$ global prompt. Here, $R_{P(i)}$ can carry the single global prompt, or both global and boundary prompts, and can include further units $R_{P(n,i)}$ intended for a similar role. 

These complement each other regarding the order in which they are assembled: $G_{sp}$ conveys role, background, attitude and style directives for $LLM_X$, whereas $B_{sp(i)}$ forwards guidelines to how $LLM_X$ shall assemble the answer such as work process, topic structure, stylesheet and generic codeblock format as applicable. Then, the user can compose/import several $'n'$ policy units $R_{P(n,i)}$ to complement the policy component $R_{P(i)}$ for a single call to $LLM_X$.

\subsubsection{Imported references} \label{sec:references}

One can import reference context units to provide basis for the workings of $LLM_X$ function, to guide the process at a closer approximation to the intended reasoning mechanism for the query. These context units will work as reference basis to support the reasoning process by the $LLM_X$ function to process the core question. In an ordered context workload, the references are intended for placement between the policy units and the question units. Then, the professional user can compose the following string units:
\begin{itemize}
    \item $P_{P(i)} \,\, -\,\, $ \textit{\_reference\_component\_}:  \\  the reference component added to the context intake towards $LLM_X$.
    \begin{itemize}
        \item $P_{p(i)} \,\, -\,\, $\textit{\_single\_reference\_}\,: \\ 
        an individual reference unit that will be directly added to the context,
        \item $P_{P(p,i)} \,\, -\,\, $\textit{\_several\_references\_}\,: \\ 
        several $'p'$ reference units that will be assembled together and added to the context,
    \end{itemize}
\end{itemize}

The user will first import the context unit(s) he expects to forward to $LLM_X$; here, he can add a single $P_{p(i)}$ context unit, or can add several $P_{P(p,i)}$ reference units to the context workload. Here, $P_{P(i)} \, (I)$ considers the import of a single reference unit $P_{p(i)}$, whereas $P_{P(i)} \, (II)$ considers the import of several reference units $P_{P(p,i)}$. The assembled context is then sent to $LLM_X$. 
\vspace{-12pt}

{\large\begin{equation}\label{eq:eq6a}
	P_{P(i)} = \underbrace{P_{p(i)}}_{\text{single}} \,\,\,\, \bigg| \,\,\,\, P_{P(i)} = \Big|\Big|^{P}_{p=1} \space \underbrace{P_{P(p,i)}}_{\text{ref. unit}} \,\,\, \bigg| \,\,\, \forall \,\,{p} \in\mathbb{N}
\end{equation}}

These context units can be individually composed/imported by the user at anytime before the call. Here, an imported context unit $P_{p(i)}$ serves as reference towards the call $'i'$. to $LLM_X$, according to the equation \ref{eq:eq6a} above. At the same time, the reference component can include several $'p'$ context units towards the call $'i'$, from the first to the final $'P_{th}'$ reference, each making a context unit $P_{P(p,i)}$ to $LLM_X$.

The user makes all considerations of purpose and ordering about importing the reference units, one or multiple at a time.  The effects of the formulation of the reference component $P_{P(i)}$ are only constrained by the limit within the LLM context window as defined by the selected $LLM_X$ function; the context window limit affects the functionality considering references the answer $A_{M(i)}$, with effects explained in the section \ref{sec:attention} of this paper. 

\subsubsection{Core question operations} \label{sec:core-question}
\begin{figure}[htbp]
\centering
    \begin{tikzpicture}[
            lifeline/.style={dashed, thick},
            actbar/.style={draw, fill=gray!20, thick, minimum width=8pt},
            msg/.style={-{Latex[length=6pt]}, thick},
            actor/.style={draw, fill=white, thick, minimum width=40pt, minimum height=20pt, align=center, font=\small},
        ]
        \def\lA{0.5}    
        \def\lB{3.0}    
        \def\lC{5.0}    
        \def\lD{7.0}   
        \def\lDd{9.5}
        \def\lE{12.0}
        \def\topY{-0.5}
        \def\botY{-10}
        \def\tI{-2}
        \def\tII{-2.75}
        \def\tIII{-3.5}
        \def\tIV{-4.25}
        \def\tV{-5.0}
        \def\tVI{-5.75}
        \def\tVII{-6.5}
        \def\tVIII{-7.25}
        \def\tIX{-8.0}
        \def\tX{-8.75}
        \node[actor] at (\lA, \topY) {User};
        \node[actor] at (\lB, \topY) {$Q_{p(i)}$};
        \node[actor] at (\lC, \topY) {$O_{v(i)}$};
        \node[actor] at (\lD, \topY) {$Q_{v(m,i)}$};
        \node[actor] at (\lDd, \topY) {$Q_{P(i)}$};
        \node[actor] at (\lE, \topY) {$LLM_X$};
        \draw[lifeline] (\lA, \topY-0.5) -- (\lA, {\botY+0.9});
        \draw[lifeline] (\lB, \topY-0.5) -- (\lB, {\botY+0.9});
        \draw[lifeline] (\lC, \topY-0.5) -- (\lC, {\botY+0.9});
        \draw[lifeline] (\lD, \topY-0.5) -- (\lD, {\botY+0.9});
        \draw[lifeline] (\lDd, \topY-0.5) -- (\lDd, {\botY+0.9});
        \draw[lifeline] (\lE, \topY-0.5) -- (\lE, {\botY+0.9});
        \foreach \lbl/\y in {
            {$compose_{(i)}$}/\tI,
            {$Q_{p(i)} \rightarrow$}/\tII,
            {$compose_{(i)}$}/\tV,
            {$O_{v(i)} \rightarrow$}/\tVI,
            {$compose_{(q,i)}$}/\tVII,
            {$|| \, Q_{v(q,i)} \rightarrow$}/\tVIII,
            {$|| \, Q_{v(Q,i)} \rightarrow$}/\tIX,
            {$set_{Q_{v(q,i)}}$}/\tIV,
            {$Q_{P(i)}\,(II)>$}/\tX,
            {$Q_{P(i)}\,(I)>$}/\tIII,
            }{
            \node[font=\footnotesize\itshape, anchor=east] at (0.5, \y) {\lbl};
        }
        \foreach \y in {\tI, \tII, \tIII, \tIV, \tV, \tVI, \tVII, \tVIII, \tIX, \tX}{
            \draw[gray!25, thin] (0.5, \y) -- (12, \y);
        }
        \draw[->>] (\lA, \tI) -- node[above, font=\scriptsize]{Set $Q_{p(i)}$} (\lB, \tI);
        \draw[->>] (\lA, \tV) -- node[above, fill=white, text opacity=1, font=\scriptsize]{Set vectoring operator $O_{v(i)}$} (\lC, \tV);
        \draw[->>] (\lA, \tVII) -- node[above, fill=white, text opacity=1, font=\scriptsize]{Set prompt vectors $Q_{v(q,i)}$} (\lD, \tVII);
        \draw[msg] (\lB, \tII) -- node[above, fill=white, text opacity=1, font=\scriptsize]{Query statement to $Q_{P(i)}$} (\lDd, \tII);
        \draw[msg] (\lC, \tVI) -- node[above, fill=white, text opacity=1, font=\scriptsize]{Vectoring operator $O_{v(i)}$} (\lDd, \tVI);
        \draw[msg] (\lD, \tVIII) -- node[above, fill=white, text opacity=1, font=\scriptsize]{Set $|| \, Qv_{(q,i)}$} (\lDd, \tVIII);
        \draw[msg] (\lD, \tIX) -- node[above, fill=white, text opacity=1, font=\scriptsize]{Set $|| \, Qv_{(Q,i)}$} (\lDd, \tIX);
        \draw[->>] (\lA, \tIV) -- node[above, fill=white, text opacity=1, font=\scriptsize]{Set vectoring mode as combo with $Q_{v(q,i)}$} (\lDd, \tIV);
        \node[fill=white, text opacity=1, font=\normalsize, anchor=east] at (\lDd, \tIII) {$Q_{c(i)}\, \in \, Q_{P(i)}$};
        \draw[msg] (\lDd, \tIII) -- node[above, font=\scriptsize]{eq. (9, left)} (\lE, \tIII);
        \node[fill=white, text opacity=1, font=\normalsize, anchor=east] at (\lDd, \tX) {$||^{Q}_{q=1} \,\,  Q_{c(i)}\, ||\, O_{v(i)}\, ||\, Q_{v(q,i)}=Q_{P(i)}$};
        \draw[msg] (\lDd, \tX) -- node[above, font=\scriptsize]{eq. (9, right)} (\lE, \tX);
        \draw[thick, rounded corners=4pt]
            (-1.6, 0.65) rectangle (13.6, \botY+0.3);
        \node[font=\small\bfseries, anchor=north west] at (-1.6, 0.65)
            {Question workflow: $Q_{P(i)}$};
    \end{tikzpicture}
    \caption{Assembly operations to core question $Q_{P(i)}$ for a single $LLM_X$ call.}
    \label{fig:question-diagram}
\end{figure}
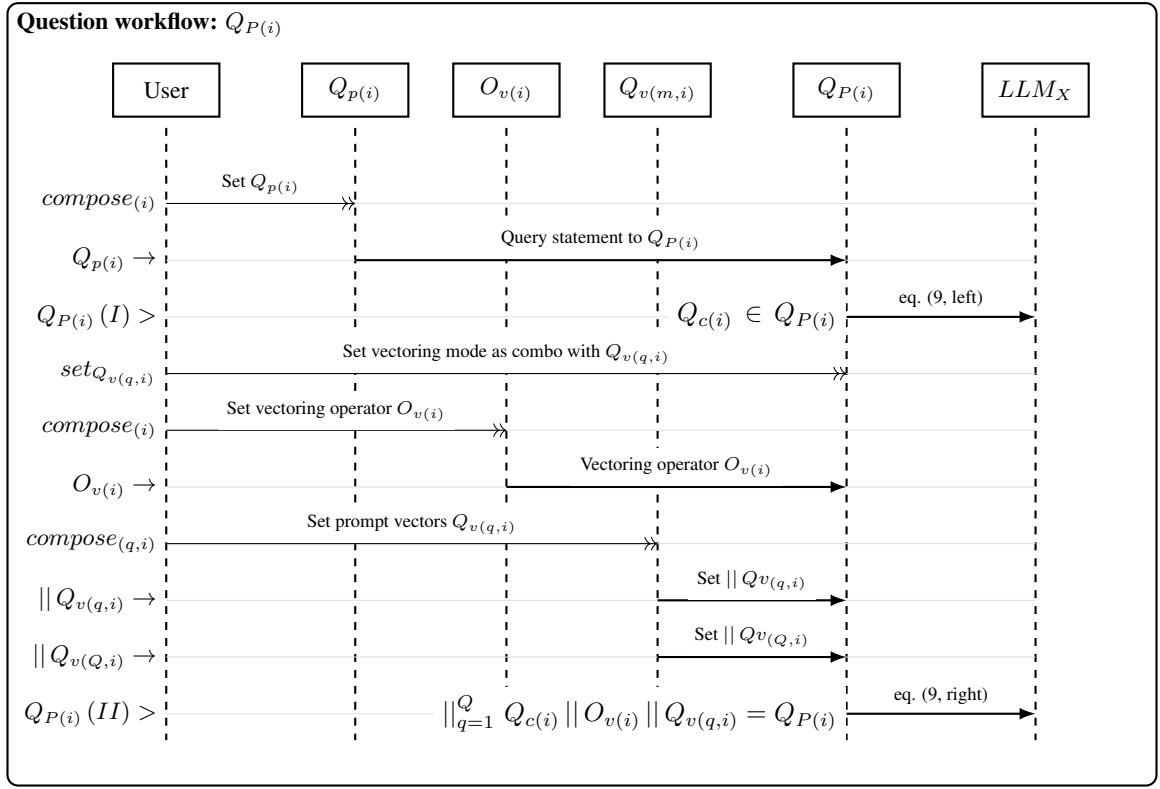

The context assembly to $LLM_X$ calls involves the crafting of a question for each request, in the form of a query statement that is intended to trigger attention by the model. This is a key component to steer the internals of the selected $LLM_X$ function, as it defines the object of inquiry and thus the focus of the process. These context units, composed through the process in the Figure \ref{fig:question-diagram} will work as reference basis to support the reasoning process by the $LLM_X$ function to process the core question. 

Then, the professional user can compose the following string units:
\begin{itemize}
    \item $Q_{P(i)}:$ \textit{\_core\_question\_}  \\ the question statement as elected by the user to call $LLM_X$.
    \begin{itemize}
        \item $Q_{p(i)}:$ \textit{\_query\_statement\_} \\ the query statement written by the user towards its intent for $LLM_X$,
        \item $O_{v(i)}:$ \textit{\_vectoring\_operator\_} \\ a relationship  operator clause  to steer $Q_{c(i)}$ onto specifics,
        \item $Q_{v(m,i)}:$ \textit{\_prompt\_vector\_} \\ an aspect clause that adds a specific to $Q_{c(i)}$ that $LLM_X$ shall process,
    \end{itemize}
\end{itemize}

The user will first compose the question he expects to forward to $LLM_X$; here, he can add the query statement $Q_{p(i)}$ alone, or can set a prompt vectoring  operation.  with a vectoring operator $O_{v(i)}$ with $'m'$ prompt vectors to request $LLM_X$ to answer the query statement in several specific aspects, even with the ability to associate $Q_{x(i,m,k_m)}$ exemplars to each vector. Once being set about the core question $Q_{P(i)}$ before calling $LLM_X$, the user can proceed to perform the query.

The user needs to compose the core question $Q_{P(i)}$ towards calling $LLM_X$. Here, the user can compose a single query statement $Q_{c(i)}$ as in display by Equation \ref{eq:eq10a}. For that purpose, the query statement will contain guidewords such as \textit{'what'}, \textit{'where'}, \textit{'how'}, or elaborated requests such as with including \textit{'please explain'} or \textit{'I need to know'} clauses.  These words play the role of attention-triggers telling $LLM_X$ to focus on the specific aspect of the query statement, and to provide a response that is compliant with the intent of the user. 
\vspace{-12pt}

{\large\begin{equation}\label{eq:eq10a}
	Q_{P(i)} = \underbrace{Q_{c(i)}}_{\text{query}} \quad \bigg| \quad Q_{P(i)} = \big|\big|_{q=1}^Q \, \underbrace{Q_{c(i)}}_{\text{query}} \,\, \big|\big| \,\, \{\space \, \underbrace{O_{v(i)}}_{\text{operator}} \, \big|\big| \, \underbrace{Q_{v(q,i)}}_{\text{vectors}}\space \,\} \quad \Big| \quad \forall \,\, q \in \mathbb{N}
\end{equation}}
\vspace{6pt}

At the same time, the user may figure the query statement can be enhanced by \textit{prompt vectors}. Here, the question $Q_{P(i)}$ as shown by equation \ref{eq:eq10a} aggregates the clauses for $Q_{c(i)}$, and then the  $O_{v(i)}$, repeated times to each $Q_{v(i,m)}$ elements, and then $Q_{P(i)}$ will ask the $LLM_X$ function to provide an aggregate answer considering all vectors according to Equation \ref{eq:eq10a}, under guidance by the rulework first provided in $R_{P(i)}$ and with reference to imported context units within $P_{P(i)}$. 

The prompt vectoring formulation in Equation \ref{eq:eq10a} applies to the assembly of a single request to $LLM_X$ function, and enables comprehensive responses upon the capability of individual $LLM_X$ functions. The individual answer provided by $LLM_X$ will address all single specifics as defined in prompt vectors, because the query statement $Q_{c(i)}$ and the operator $O_{v(i)}$ are replicated at all times along each prompt vector in the core question component.

\subsection{Context assembly operations} \label{sec:context-assembly}

The assembly of context towards engineering tasks, from system design context definition to model-building and implementation, requires flexibility and modularity in designing and handling systems information as context input for use with large language models, through performing the context operations formalized in section \ref{sec:operations}. Table~\ref{tab:cx-modules} lists the role-focused context units in the context assembly process as defined in the previous section, and the intent for each.  

\begin{table}[htbp]
    \centering
    \caption{Context modules and units used in the assembly of the context workload forwarded to the LLM.}
    \label{tab:cx-modules}
    \begin{tabular}{L{1.5cm} L{3.25cm} C{2cm} S{5cm}}
        \toprule
         & \textbf{Context unit} & \normalsize\textbf{Operation} & \normalsize\textbf{Intent} \\
        \midrule
        \textit{\textbf{\large R\textsubscript{P}}} & \textbf{Policy rulework} & section \ref{sec:policy} & \\
        $\rightarrow G_{sp}$ & \makecell[lt]{Global prompt}  & &
        \makecell[lt]{Defines engineering role, epistemic rules, \\ and non-hallucination constraints}  \\
        $\rightarrow B_{sp}$ & \makecell[lt]{Boundary prompt} & &
        \makecell[lt]{Defines system scope to its \\ intended application and characteristics.} \\
        \midrule
        \textit{\textbf{\large P\textsubscript{P}}} & \textbf{Reference priors} & section \ref{sec:references} & \\
        $\rightarrow P_{P}$ & Full References & &
        \makecell[lt]{Provides references with descriptions \\ and state-of-the-art technology.} \\
        $\rightarrow P_p$ & Other references & &
        \makecell[lt]{Extracts condensed system and design \\ constraints from imported sources.} \\
        \midrule
        \textit{\textbf{\large Q\textsubscript{P}}} & \textbf{Core question} & section \ref{sec:core-question} & \\
        $\rightarrow Q_p$ & Query statement  & & Sets primary architectural inquiry. \\
        $\rightarrow O_v$ & Vectoring operator & &
        \makecell[lt]{Field: Sets relationships between partial \\ aspects and the main inquiry.} \\
        $\rightarrow Q_v$ & Prompt Vectors & &
        \makecell[lt]{Field(s): Defines specific aspects that \\ the task shall consider.}\\
        \midrule
        \textit{\textbf{\large C\textsubscript{Q}}} $\Rightarrow$  & \textbf{Context workload} & section \ref{sec:context-assembly} &
        \makecell[lt]{Context workload assembled \\ from context operations.} \\
        \midrule
        \textit{\textbf{\large LLM\textsubscript{X}}} & \textbf{Large language model} & section \ref{sec:param-llm} & \\
        $\Rightarrow A_M$ & Answer yield & &
        \makecell[lt]{LLM-generated content that considers the \\ information in the context workload.}  \\
        \bottomrule
    \end{tabular}
\end{table}

This setting of different context units with basis on roles enables the assembly of a context workload $C_Q$ to be forwarded to the $LLM_X$ function with mind to a certain intent towards a work product that shall be embodied by the means of the answer yield $A_M$. However, the effectiveness of the context workload $C_Q$ depends on the assembly of the context units, which requires understanding the dependencies between them. 

The reason for this understanding lies in the dependency relations between context units as expressed by the context operation definitions from section \ref{sec:operations}: the joining and assembly of context unities takes place by means of concatenation; the \textit{transformer} mechanism within large laguage models \cite{vaswani_attention_2017} is such that the order of the context units in the workload matters, and the dependencies between them are relevant to the answer yield $A_M$. 

This means dependencies between context units must respected when assembling the context workload $C_Q$ Table~\ref{tab:dsm} presents a Design Structure Matrix (DSM) capturing dependency relations between context units. The context assembly takes place before the call to $LLM_X$, from which the internals of the model generate the answer.

\begin{table}[htbp]
    \centering
    \caption{DSM Representing context assembly, $LLM_X$ event horizon and answer yield with memory.}
    \label{tab:dsm}
    {\renewcommand{\arraystretch}{1.5}
        \begin{tabular}{lccccccccccccc}
            \toprule
            &  $G_{sp}$ & $B_{sp}$ & $R_P$ & $P_r$ & $P_r$ & $P_P$ & \textbf{$Q_c$} & $O_v$ & $Q_v$ & $Q_P$ & $C_Q$ & $LLM_X$ & $A_M$ \\
            \midrule
            $G_{sp}$ & -- &   &   &   &   &   &   &   &   &   &   &  | \\
            $B_{sp}$ & $\bullet$ & -- &   &   &   &   &   &   &   &   &   &  | \\
            $R_P$ & $\bullet$ & $\bullet$ & -- &   &   &   &   &   &   &   &   &  | \\
            $P_r$ &  &  &  & -- &  &   &   &   &   &   &   & |  \\
            $P_r$ &  &  &  & $\bullet$ & -- &  &   &   &   &   &   & |  \\
            $P_P$ &  &  &  & $\bullet$ & $\bullet$ & -- &  &   &   &   &   &  | \\
            $Q_p$ &  &  &  &   &   &   & -- &  &   &   &   & |  \\
            $O_v$ &  &  &  &   &   &   & $\bullet$ & -- &  &   &   & |  \\
            $Q_v$ &  &  &  &   &   &   & $\bullet$ & $\bullet$ & -- &  &   & |  \\
            $Q_P$ &  &  &  &   &   &   & $\bullet$ & $\bullet$ & $\bullet$ & -- &  &  | \\
            $C_Q$ & $\rightarrow$ & $\rightarrow$ & $||$ & $\rightarrow$ & $\rightarrow$ & $||$ & $\rightarrow$ & $\rightarrow$ & $\rightarrow$ & $||$ & -- &  | \\
            \midrule
            $LLM_X$ &  &   &   &   &   &   &   &   &   & $\rightarrow$  & $\bullet$ & -- \\
            \midrule
            $A_M$ &  &   &   &   &   &   &  &   &   &  & $\rightarrow$ & $\bullet$ & ($\bullet$)\\
            \bottomrule
            \multicolumn{14}{l}{\scriptsize\parbox{0.8\linewidth}{\raggedright
            LEGEND:\\
            $\bullet$ Row items succeeding column items.\quad \,\,
            $\rightarrow$ Concatenation of items onto modules.\quad \,\,
            $||$ Operations on context modules.\par}} \\
        \end{tabular}
        }
        \vspace{4pt}
\end{table}

The DSM displays the correspondence and the resulting order of context units, from the upstream policy units, through the imported references placed in between, and the core question units downstream. Three principles steer this ordering sequence: (i) the order of context units in the workload matters due to the transformer mechanism; (ii) the dependencies between context units are relevant to the answer yield $A_M$; and (iii) the memory of $LLM_X$ ends up prioritizing the beginning and the end of the context workload.

A person in the role of \textit{systems design engineer} curates and composes/imports the relevant context units, thus assembling the context workload, and then requests $LLM_X$. The chosen LLM will process the context workload $C_Q$ with the parameters set in section \ref{sec:param-llm} and yield its answer $A_M$ under directives and guidelines set within the policy rulework module $R_P$, with basis on the $P_P$ imported references, and supported by the exemplars set within $Q_P$. 

\section{Case study}
This section demonstrates the outputs of system configuration requests enabled by the context operations formally defined in section \ref{sec:operations}. This study focuses the assembly of context units to support the processing of $LLM_X$ functions at building component architecture models of complex systems. To demonstrate how context operations work onto supporting these functions, this paper involves a modelling case  regarding the component architecture of a hybrid SAR UAV, which is hereby denoted as \textit{Suthern-Cross} as in display by Table \ref{tab:project-description}. 

{\renewcommand{\arraystretch}{1.25}
\begin{table}[htbp]
    \centering
    \caption{Context operations demonstration assignments regarding systems characteristics}
    \label{tab:project-description}
    \begin{tabular}{L{2.5cm} S{6.5cm} C{2.5cm}}
        \toprule
         & {\normalsize\textbf{Heavy SAR UAV - Suthern-Cross}} & \multirow{5}{*}{\includegraphics[width=2.4cm]{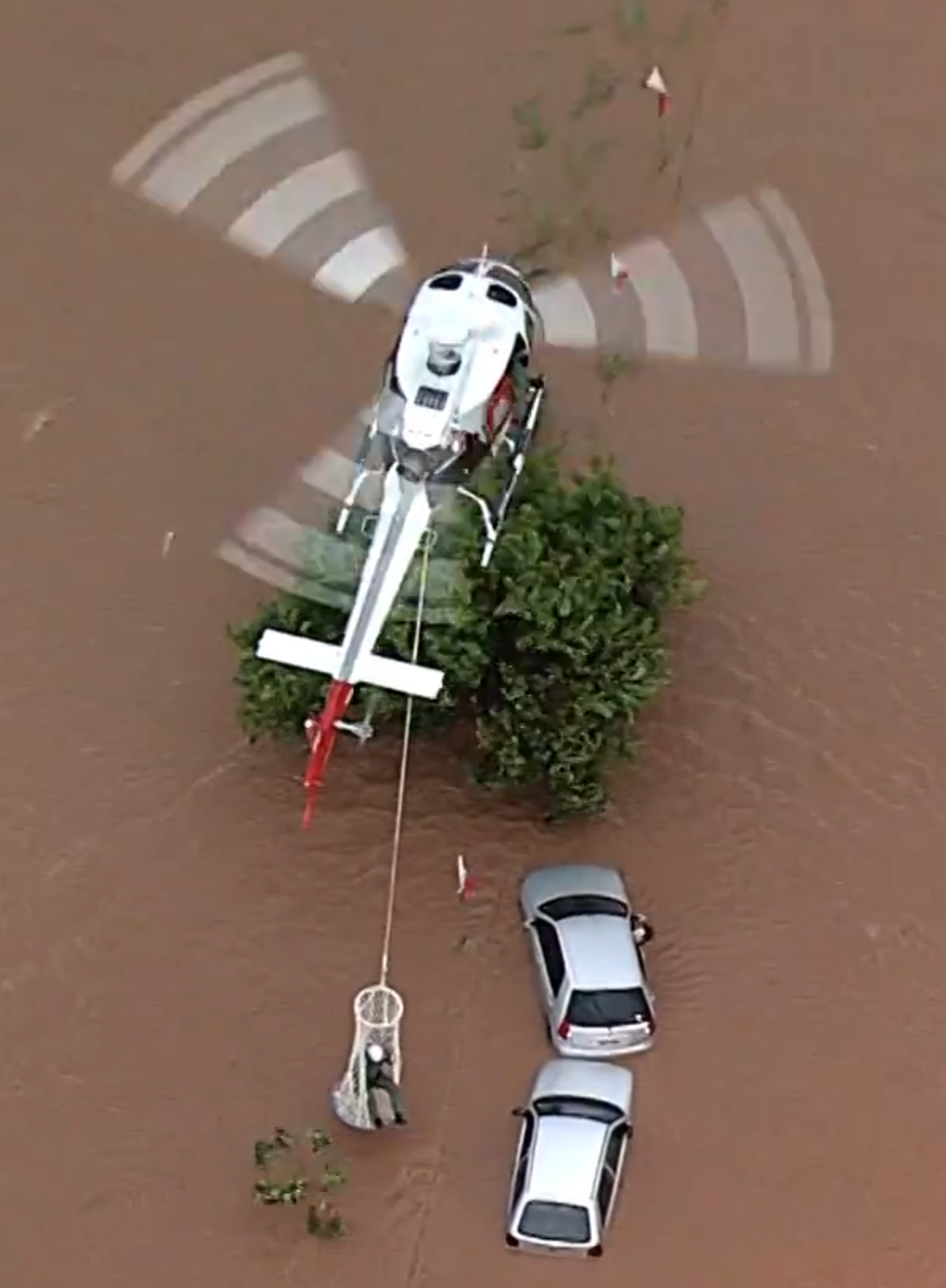}} \\
        \cmidrule{1-2}
        \textbf{\makecell[lt]{Assignment}} &
        Operation: Search-and-rescue operations to lift and recover victims from flash flood to medical care. \\
        \textbf{\makecell[lt]{Key\\ requirements}} & Station-keeping on hover, limited downwash over victims, low noise footprint, aircraft regulatory compliance. \\
        \textbf{\makecell[lt]{Key\\ constraints}} & Heavy rainfall at location, aggressive wind and gusts, timespan to victim acquisition, energy and power budget. \\
        \textbf{\makecell[lt]{LLM\\ Assignment}} & Help functional decomposition and architecture design under professional supervision towards defining  system/concept architecture. \\
        \bottomrule
    \end{tabular}
    \\[2pt]
    {\scriptsize\centering
    \textbf{Picture source:} \url{https://youtu.be/QVyptRhdDgA?si=ygtpMefwQXCHfO54&t=84}, accessed on 2026-06-05.\par}
\end{table}
}

Table \ref{tab:project-description} includes a depiction of an as-current Search-and-Rescue operation over flooded area with an helicopter - the SAR UAV is expected to have similar capability of recovering victims. The architecture modelling task is defined with the goal of creating a functional and operational system, starting from a system scope of required functionality. The focus of this case study is the leveraging of the modelling-as-code paradigm, where the modelling language is used to generate a model from the answer yield $A_M$ provided by $LLM_X$.

\subsection{Context workload treatments}\label{sec:work-treatments}

The context treatments involve selective use of context units onto requesting $LLM_X$ for the generation of the intended work product. The context units within the treatments follow the definitions set in section \ref{sec:operations}; the role of individual context units towards the assembly of input to $LLM_X$ is considered upon composing and importing information content, as in display by Table \ref{tab:context-workload-treatment}.  

{\renewcommand{\arraystretch}{1.25}
\begin{table}[htbp]
    \centering
    \caption{Context workload treatments.}
    \label{tab:context-workload-treatment}
    \footnotesize
    \begin{tabular}{L{1.2cm} L{1.95cm} L{1.95cm} L{1.95cm} L{1.95cm} L{1.95cm}}
        \toprule
         & \textbf{$G_{sp}$ Global} & \textbf{$B_{sp}$ Boundary} & \textbf{$P_P$ Reference} & \textbf{$Q_c$ Query} & \textbf{$O_v\space||\space Q_v$Vectors} \\
        \midrule
        \textbf{Content} & \makecell[lt]{Role directives,\\mission scenario,\\Validation} & \makecell[lt]{Modelling scope,\\principles, rules,\\requirements} & \makecell[lt]{Generated\\benchmarking\\analysis.} & \makecell[lt]{Work product\\request, and\\requirements.} & \makecell[lt]{$\mu$-Template\\subsystem\\exemplars.}  \\
        \addlinespace[3pt]
        \midrule
        \makecell[lt]{\textbf{T1}}  &  &  &  & \makecell[lt]{$\bullet\bullet\bullet$\textbf{>}\\ \textnormal{127 Tk}} &  \\
        \makecell[lt]{\textbf{T2}}  & $\bullet$\textbf{++>} & $\bullet$\textbf{++>} &  & \makecell[lt]{$\bullet\bullet\bullet$\textbf{>}\\ \textnormal{2485 Tk}} &  \\
        \makecell[lt]{\textbf{T3}}  & $\bullet$\textbf{++>} & $\bullet$\textbf{++>} &  & \makecell[lt]{$\bullet\bullet\bullet$\textbf{>}} & \makecell[lt]{$\bullet$\textbf{++>}\\ \textnormal{7645 Tk}} \\
        \makecell[lt]{\textbf{T4}}  & $\bullet$\textbf{++>} & $\bullet$\textbf{++>} & $\bullet$\textbf{++>} & $\bullet\bullet\bullet$\textbf{>} & \makecell[lt]{$\bullet$\textbf{++>}\\ \textnormal{20555 Tk}} \\
        \addlinespace[6pt]
        \textbf{Input} & \makecell[lt]{Typed/pasted\\prose \& topics,\\saveable.} & \makecell[lt]{Typed/pasted\\prose \& topics,\\saveable.} & \makecell[lt]{Pasted source\\text import\\or DOCX, PDF} & \makecell[lt]{Typed/pasted\\or imported\\prose \& topics} & \makecell[lt]{Paste into\\or import} \\ 
        \addlinespace[3pt]
        \bottomrule
    \end{tabular}
\end{table}
}

The treatments are designed to be cumulative, where each treatment adds more context units to the workload\footnote{The context intake length in tokens [Tk] is the workload that utilizes context length capacity in each of the models called. Table \ref{tab:llm-setup-benchmarking} displays model characteristics regarding context length and answer length capacities for reference.} for $LLM_X$ to process. The cumulation of treatments is such that the first treatment T1 only includes the query statement $Q_c$; the second treatment T2 adds the global and boundary prompts $G_{sp}$ and $B_{sp}$; the third treatment T3 adds the prompt vectors $O_v$ and $Q_v$; and the fourth treatment T4 adds the reference priors $P_P$. 

The treatments are designed to evaluate the effects of different combinations of context units on the answer yield $A_M$ from $LLM_X$, which are assembled with basis on the context operations defined in section \ref{sec:operations}.  For that purpose, the experiment is based on examples of using \verb+plantUML+ as modelling language, considering LLM-aided modelling technique, such as by \cite{camara_Assessment_2023} and \cite{Krus_LLMSAerospaceICAS_2024}. 

\subsection{Model use treatments}\label{sec:llm-treatments}

Besides the four context treatments, the experiment also includes five $LLM_X$ model treatments. The model assortment for this case study is designed to evaluate the effects of model characteristics on the answer yield $A_M$ from $LLM_X$ receiving the same context workload. The LLM assortment is designed to evaluate the effects of model characteristics on the answer yield $A_M$ from $LLM_X$ receiving the same context workload. 

The LLM assortment was defined with mind to considering processing resource: one model running on local GPU\footnote{The local model runs with a 4Gb Nvidia Quadro T1000 GPU and a 2.7-4.0 GHz 12-core Intel Core-i7 CPU with 16Gb memory.}, two models from the cloud\footnote{Cloud bandwidth intake takes place at 50 Mbps and answer yield takes place at 500 Mbps.} that can be run on high-end local desktop GPUs to 48Gb RAM, and two frontier models to be run from the cloud. Table \ref{tab:llm-setup-benchmarking} displays the specifications, their descriptions and their units. The API interface forwards the assembled context workloads to the selected $LLM_X$ models. 


{\renewcommand{\arraystretch}{1.25}
\begin{table}[htbp]
    \centering
    \caption{Model structure metrics - characterization of models.}
    \label{tab:llm-setup-benchmarking}
    \footnotesize
    \begin{tabular}{L{2cm} L{2cm} L{2cm} L{2cm} L{2cm} L{2cm}}
        \toprule
        & M1 & M2 & M3 & M4 & M5 \\
        \cmidrule(lr){2-6}
        & \makecell[lt]{\textbf{Qwen3}\\ \textbf{8b}} & \makecell[lt]{\textbf{ChatGPT}\\ \textbf{OSS-20b}} & \makecell[lt]{\textbf{Nemotron3}\\ \textbf{super-120b-a12b}} & \makecell[lt]{\textbf{Kimi}\\ \textbf{K2.5}} & \makecell[lt]{\textbf{Claude}\\ \textbf{Sonnet4.6}} \\
        \cmidrule(lr){2-6}
        & \scriptsize{\cite{qwen3}} & \scriptsize{\cite{openai2025gptoss}} & \scriptsize{\cite{nemotron3super}} & \scriptsize{\cite{ollama_kimi_k25_2026}} & \scriptsize{\cite{sonnet46}} \\
        \midrule
        \textbf{Resource} & Local/GPU & Cloud/Nvidia & Cloud/Nvidia & Cloud/Ollama & Cloud/Anthropic \\
        \makecell[lt]{\textbf{Architecture}\\ \textnormal{[Param.]}} & \makecell[lt]{Distilled\\ $8\times10^9$} & \makecell[lt]{MoE\\ $2\times10^{10}$} & \makecell[lt]{MoE\\ $1,2\times10^{11}$} & \makecell[lt]{MoE\\ $1\times10^{12}$} & Undisclosed  \\
        \makecell[lt]{\textbf{Context} \textnormal{[Tk]}} & $3,27\times10^4$ & $1,28\times10^5$ & $1,00\times10^6$ & $2,56\times10^5$ & $1,00\times10^6$ \\
        \makecell[lt]{\textbf{Answer} \textnormal{[Tk]}}  & $4,09\times10^3$ & $3,27\times10^4$ & $6,40\times10^4$ & $6,40\times10^4$ & $6,40\times10^4$ \\ 
        \bottomrule 
    \end{tabular}
\end{table}
}

The following characteristics affect modelling performance: (i) local hosting is limited to the memory room in the local machine; (ii) a higher scale of parameter count usually enables more complex and targeted processing; (iii) a longer context length capacity means the maximum number of tokens that can be processed by an individual LLM; and (iv) the answer length limit is less about capability, and more about policy - especially with higher-end models.

\subsection{Modelling corpus requirements}\label{sec:model-corpus-requirements}

The case study proceeds with the assumption that the modelling response can provide a work product the interpreter is able to render. Context workload components involve the same content set by context composition treatment regardless of the model called, so the resulting models can be compared across treatments and $LLM_X$ models to evaluate the role of context operations and model scale. 

In the modelling-as-code approach, the context workload conveys to $LLM_X$ a request for a system model intended to represent the intended system architecture with sufficient detail and within a configuration that complies with required functionality. To enable the proper validation of the model outputs in the corpus of generated models across treatments, two sets of modelling requirements are defined: 

\begin{enumerate}
    \item[a)] \verb+plantUML+ syntax convention; and 
    \item[b)] architecture requirements defined in Table \ref{tab:plantuml-model-requirements}.
\end{enumerate}

Considering the context workloads from Table \ref{tab:context-workload-treatment} submitted to the $LLM_X$ models from Table \ref{tab:llm-setup-benchmarking} include requests for \verb+plantUML+ cobeblocks as part of the answers $A_M$, the evaluation of the capability by the models to process context workloads takes place through the manual checking and validation of the codeblocks inside the answer files. For that purpose, the model outputs in conversation codeblocks are exported onto \verb+plantUML+ model code files, to be verified and corrected individually in the following steps:
\begin{enumerate}
    \item Verify the model $m_0 \in A_M$ in the answer codeblock from $LLM_X$ with, duplicating it onto a $m_1 = m_0$ \texttt{plantUML} model file for syntax review with the interpreter;
    \item verify the $m_1$ model file for syntax errors and architecture requirements, iterate corrections until the model is fully correct, then save the corrected model $m_2 \neq m_1$ to a model file with \texttt{Corr1} suffix;
    \item Verify the corrections made to $m_2$ on architectural requirements, amend it to comply with the requirements and save the corrected model $m_3 \neq m_2$ to a model file with \texttt{Corr2} suffix.
\end{enumerate}

Then, the evaluation of the resulting corpus involves the review and validation of model-as-code units with metrics conveying quality requirements from Table \ref{tab:plantuml-model-requirements}.

{\renewcommand{\arraystretch}{1.25}
\begin{table}[htbp]
    \centering
    \caption{Modelling requirements for system architecture models.}
    \label{tab:plantuml-model-requirements}
    {\fontsize{8pt}{9pt}\selectfont
    \begin{tabular}{L{1.5cm} L{2.5cm} L{2.5cm} L{2.5cm} L{2.5cm} L{2.5cm}}
        \toprule
            & \multicolumn{5}{l}{\normalsize\textbf{plantUML syntax rules}} \\
        \addlinespace[5pt]
        \normalsize{\textbf{\makecell[lt]{Model\\syntax}}} & \makecell[lt]{Unique \textbf{component}\\\texttt{name\_def} definitions\\in entity syntaxes:\\{\texttt{[component\_name]}},\\or {\texttt{component "Name"}},\\with unique {\texttt{as alias}}.} & \makecell[lt]{Unique \textbf{port}\\ \texttt{name\_def} definitions\\in entity syntaxes:\\ port PT within element,\\ or {\texttt{port "PT"}} under\\component \texttt{name\_def}.} & \makecell[lt]{Single-pair \textbf{flows} with\\{\texttt{cp1 - cp2:src-tgt}}\\source-sink definition\\and matching \textbf{port} links\\ with correct arrows to\\flow sink.} & \makecell[lt]{\textbf{Flow} direction signs\\and \texttt{[format]} params\\between dashes and\\no stray \textbf{component}\\or \textbf{port} {\texttt{name\_defs}}\\ outside entity syntaxes.} & \makecell[lt]{Traceable, consistent\\{\textbf{component}} and {\textbf{port}}\\\texttt{name\_def} assignment\\throughout \textbf{component}\\entity definitions to\\\textbf{flow} statements.} \\
        \addlinespace[5pt]
        \midrule
            & \multicolumn{5}{l}{\normalsize\textbf{Component architecture requirements}} \\        
        \normalsize{\textbf{\makecell[lt]{Model\\elements}}} & \makecell[lt]{Single energy source\\with fuel specification\\and power plant with\\ controls, power lines,\\drivetrain and end-\\effect components.} & \makecell[lt]{Power take-off and\\distribution with heat\\exchange components,\\end-effects linked to\\structural elements\\such as airframe/pylons.} & \makecell[lt]{Safety and payload\\systems including\\parachute, hoist, winch,\\cabling, load cell\\and control with\\rescue equipment.} & \makecell[lt]{Flight control systems\\including sensory and\\processing components,\\along connectivity and\\controls to ancillary \&\\end-effect actuators.}  & \makecell[lt]{Flight support and,\\and onboard mission\\ controls and data\\processing with\\communication systems\\and protocols.} \\
        \addlinespace[5pt]
        \normalsize{\textbf{\makecell[lt]{Model\\ports/flows}}} & \makecell[lt]{\textbf{Component} \texttt{name\_def}\\shall have one or more\\individual \texttt{name\_def}\\ \textbf{ports} within or under\\\texttt{name\_def} statement.} & \makecell[lt]{\texttt{name\_def} \textbf{flow} links to\\same \texttt{name\_def} \textbf{ports}\\in single component at\\source and in single\\component at sink.} & \makecell[lt]{There is no stray\\\texttt{name\_def} \textbf{component}\\and no \textbf{port/flow}\\\texttt{name\_def} mismatch\\by either \textbf{flow} end.} & \makecell[lt]{\textbf{Flow} \textit{linetype} setting\\within brackets in \textbf{flow}\\ definitions between\\single pair of source\\and sink \textbf{components}.} & \makecell[lt]{Single \textbf{flow} connections\\between connected pairs\\of \texttt{name\_def} \textbf{ports} by\\source and sink\\\texttt{name\_def} \textbf{components}.} \\
        \addlinespace[5pt]
        \normalsize{\textbf{\makecell[lt]{Model\\compliance}}} & \makecell[lt]{Power system energy\\\textbf{flows} from single\\source to countable\\end-effect \textbf{components}\\and outputs.} & \makecell[lt]{Control system \textbf{flows}\\from sensor elements\\through controller\\elements and actuator\\\textbf{components}.} & \makecell[lt]{One-way source-to-sink\\\textbf{flow} chains for control\\and power systems with\\no circular paths across\\ energy \& signal \textbf{flows}.} & \makecell[lt]{Control architecture\\across signal \textbf{flows} by\\acting \textbf{components} to\\ensure functionality\\and control modes.} & \makecell[lt]{Specific quantification\\of \textbf{components} about\\energy sources and\\end-effect assemblies\\and unique \textbf{ports}.}\\
        \addlinespace[5pt]
        \bottomrule
    \end{tabular}}
\end{table}
}

The inclusion of exported \texttt{plantUML} model files enables the verification of the outputs regarding the requirements above, on the following order: raw $\rightarrow$ syntax $\rightarrow$ architecture. A model is considered correct if it complies with the syntax and architecture requirements, and it is considered incorrect if it does not comply with either of them. The variety of models from Table \ref{tab:llm-setup-benchmarking} will determine different capabilities regarding the satisfaction of the requirements, which means each pair context-model may yield different levels of compliance with the requirements. 

\subsection{Modelling answer verification}\label{sec:modelling-verification}

LLM limitations can produce a representation with insufficient detail and lacking compliance to system requirements. Sometimes, the feeding system model codeblocks to the \texttt{plantUML} interpreter fails to render the model correctly because of syntax errors. Moreover, incorrect model representations manifest in differences between raw models from $LLM_X$ output and corrected models to syntax. In this context,  the metrics reflect findings on the architectural model about whether it misses any required element or whether it does not comply to rules. The metrics proposed for this case study come in display by Table \ref{tab:plantuml-model-metrics}.

{\renewcommand{\arraystretch}{1.25}
\begin{table}[htbp]
    \centering
    \caption{Model metrics considered for architecture validation.}
    \label{tab:plantuml-model-metrics}
    \small
    \setlength{\tabcolsep}{4pt}
    \begin{tabular}{>{\raggedright\arraybackslash}p{1.8cm} *{5}{>{\raggedright\arraybackslash}p{2.4cm}}}
        \toprule
            & \multicolumn{5}{l}{\normalsize\textbf{Generative modelling outputs}} \\

        \normalsize{\textbf{Model answer}} &
        \makecell[lt]{Answer length\\{[tk] $A_{M_{len}}$}\\{$\quad \in A_M$}} &
        \makecell[lt]{Context utilization\\{[tk] $C_{L_{len}}$}\\{/[tk] $C_{M_{len}}$}} &
        \makecell[lt]{Answer time\\{[min:s] $t_A$}\\{$C_Q \rightarrow A_M$}} &
        \makecell[lt]{Model length\\{[lin] $m_{len}$}\\{$\quad \space \in A_M$}} &
        \makecell[lt]{Model type\\{[type] $m_{type}$}\\{$\in \texttt{plantUML}$}} \\
        \addlinespace[10pt]
        \midrule
            & \multicolumn{5}{l}{\normalsize\textbf{plantUML syntax rules}} \\

        \normalsize{\makecell[lt]{\textbf{Model}\\\textbf{{syntax}}}} &
        \makecell[lt]{Model versions\\{[n] $m_x$ off $A_M$}} &
        \makecell[lt]{Element syntax\\{[n] $e_{err} \notin$ [type]}\\$\in \texttt{plantUML}$} &
        \makecell[lt]{Flow syntax\\{[n] $f_{err} \notin$ [type]}\\$\in \texttt{plantUML}$} &
        \makecell[lt]{Port syntax\\{[n] $p_{err} \notin$ [type]}\\$\in \texttt{plantUML}$} &
        \makecell[lt]{Dupl. \texttt{name\_def}\\{$[n]\,e_a = e_b$}\\{$\ge 2x \in A_M$}} \\
        \addlinespace[10pt]
        \midrule
            & \multicolumn{5}{l}{\normalsize\textbf{Component architecture requirements}} \\
        \normalsize{\textbf{Model elements}} &
        \makecell[lt]{Element count\\{[n] $e_{x} \in A_M$}} &
        \makecell[lt]{Element miss\\{[n] $e_{x} \in D_C$}\\{$\quad \quad \space \,\, \notin A_M$}} &
        \makecell[lt]{Stray element\\{[n] $e_{x} \in D_C$}\\{$\nexists f_{x_{in}}, f_{x_{out}}$}} &
        \makecell[lt]{Module miss\\{[n] $g_{x} \in D_C$}\\{$\quad \quad \space \,\, \notin A_M$}} &
        \makecell[lt]{No end-effect\\{[n] $z_{x} \in D_C$}\\{$\quad \quad \space \,\, \notin A_M$}} \\
        \addlinespace[10pt]

        \normalsize{\textbf{Model ports/flows}} &
        \makecell[lt]{Flow count\\{[n] $\in A_M$}} &
        \makecell[lt]{Flow miss\\{[n] $f_x \in D_C$}\\{$\quad \notin A_M$}} &
        \makecell[lt]{Port dangle\\{[n] $p_x \in D_C$}\\{$\nexists f_{x_{in}}, f_{x_{out}}$}} &
        \makecell[lt]{Port count\\{[n] $\in A_M$}} &
        \makecell[lt]{Port miss\\{[n] $p_x \in D_C$}\\{$\quad \notin A_M$}} \\
        \addlinespace[10pt]

        \normalsize{\textbf{Model compliance}} &
        \makecell[lt]{} &
        \makecell[lt]{Input misses\\{[n]}\\{$f_{x_{out}} \neq p_{x_{in}}$}\\{$\quad \quad \quad \in m_x$}} &
        \makecell[lt]{Chain misses\\{[n]}\\{$\{e_x \rightarrow e_y \rightarrow e_z\}$}\\{$\quad \quad \quad \quad \notin D_C$}} &
        \makecell[lt]{Output misses\\{[n]}\\{$\,p_{x_{out}} \neq f_{x_{in}}$}\\{$\quad \quad \quad \in m_x$}} &
        \makecell[lt]{} \\
        \addlinespace[10pt]

        \bottomrule
    \end{tabular}
\end{table}
}

The model answer metrics in the first row quantify characteristics of the answer provided by $LLM_X$ to certain context workload treatment. The other metrics are designed to quantify the characteristics of the model outputs from $LLM_X$ regarding syntax and architecture requirements. These include $A_M$ processing statistics by $LLM_X$, then all model verification criteria - syntax and architecture - apply at the raw model codeblock $m_0$ exported to the model file $m_1$.

Equation \ref{eq:eq9} expresses the relationship between the context workload $C_Q$ and the answer yield $A_M$ from $LLM_X$, where the answer length $A_{M_{len}}$ is less than or equal to the context length $C_{Q_{len}}$, and the model length $m_{len}$ is less than or equal to the answer length $A_{M_{len}}$.
\vspace{-12pt}

{\large\begin{equation}\label{eq:eq9}
	 \underbrace{LLM_X}_{C_{M_{len}} \geq} \,\, \underbrace{\{C_{Q(i)}\} = A_{M(i)}}_{{C_{Q_{len}}+A_{M_{len}}}\,=\,C_{L_{len}}} \quad \,\, \bigg| \quad \underbrace{m_0 \in\,\, A_{M(i)}}_{m_{len}\, < \, A_{M_{len}}} \quad \,\, \bigg| \quad \,\, \underbrace{m_1=m_0}_{m_1\, \text{model file}}
\end{equation}}
\vspace{6pt}

Once a $m_1$ model file is available, its verification is carried out on the following basis: the syntax of the model is checked for errors, and then the architecture of the model is evaluated for compliance with the requirements. This is done with the checking the following assertions on the model: the assertion (I) in Equation \ref{eq:eq10} regards the existence of syntax errors in the model $m_1$ regarding element, flow and port definitions, and the second condition regards the existence of duplicate \texttt{name\_def} definitions for elements. 

\vspace{-6pt}

{\large\begin{equation}\label{eq:eq10}
    \begin{split}
	\text{I:}\quad m_{err} &= \{\underbrace{\exists \,\, [e_{err},\, f_{err},\, p_{err}]\, \notin \, \normalsize{\texttt{plantUML}}}_{\text{syntax errors}}\,\, \lor \,\, \underbrace{\forall \,\,  [a, b] \,\rightarrow \, e_a=e_b}_{\text{duplicate names/aliases}}\} \,\, \in \,\, m_1\\
    \end{split}
\end{equation}}
\vspace{6pt}

Then, the assertion (II) in the first line of Equation \ref{eq:eq11} regards the architecture requirements, with the first condition being about any missing element, flow, port, module or end-effect in the model and the second condition being about any stray component or dangling port in the model. Then, the assertion (III) in the second line of Equation \ref{eq:eq11} is true if it does not comply with architecture requirements, which include any input and output mismatches in the model, or any chain of elements that is not present in the model.
\vspace{-6pt}

{\large\begin{equation}\label{eq:eq11}
    \begin{split}
    \text{II:}\quad m_{miss} &= \{\underbrace{\nexists \,\, [e_{x},\, f_{x},\, p_{x},\, g_{x},\, z_{x}]\,\, \in \,\, D_C}_{\text{element, flow, port, module or end-effect\,misses}}\,\, \lor \,\, \underbrace{[e_x,\, p_x] \,\, \nexists\,\, [f_{x_{in}} \land f_{x_{out}}]}_{\text{stray component, dangling port}}\} \,\, \in \,\, m_x\\
    \\
    \text{III:}\quad m_{short} &= \{\underbrace{\exists\,\,[f_{x_{out}} \neq p_{x_{in}} \lor  p_{x_{out}} \neq f_{x_{in}}]}_{\text{Input and output mismatches}} \,\, \lor \,\, \underbrace{\exists\,\, [e_x \rightarrow e_y \rightarrow e_z]\,\, \notin \,\, D_C}_{\text{Chain misses}} \,\,\} \; \in \; m_x\\
    \end{split}
\end{equation}}
\vspace{6pt}

The verification of architectural metrics is preconditioned by correct model syntax.This means that the model-building attempts from context workload may display different inconsistencies with requirements - different context workloads entail levels of specification to modelling-as-code LLM outputs regarding those criteria. Then, Equation \ref{eq:eq12} describes the conditions under which additional model files are needed. 

{\large\begin{equation}\label{eq:eq12}
    \begin{split}
	\text{I:}&\quad \texttt{Corr1}= \bigl\{I \Rightarrow \exists \; m_2 \neq m_1\bigr\}\;\land\;\texttt{Corr2}=\bigl\{[I \land (II \lor III)] \Rightarrow \exists \; m_3 \neq m_2\bigr\}\\
    \\
    \text{II:}&\quad \texttt{Corr1}=\;\bigl\{[\neg I \land (II \lor III)] \Rightarrow \exists \; m_2 \neq m_1, \nexists \; m_3\bigr\}\\ 
    \\
    \text{III:}&\quad \nexists \; (\texttt{Corr1, Corr2}):\;\bigl\{[\neg I \land (\neg II \land \neg III)] \Rightarrow \nexists \; m_3, \; m_2\bigr\}
    \end{split}
\end{equation}}
\vspace{6pt}

The first condition in Equation \ref{eq:eq12} applies to the case where the model $m_1$ has syntax errors, which requires a new model file $m_2 \neq m_1$ to be created with corrections to syntax and architecture.  Here, the need to make additional model files $m_2$ and $m_3$ shall involve syntax and architectural refinements to satisfy the requirements; the corrections are ladder-staged ifrom $m_1$ to $m_2$ and then from $m_2$ to $m_3 \neq m_2$. When $m_1$ has correct syntax, then it undergoes architectural review of $m_2$ to determine the need for $m_3$.

The second condition in Equation \ref{eq:eq12} applies to the case where the model $m_1$ has correct syntax, but it does not comply with architecture requirements, which requires a new model file $m_2 \neq m_1$ to be created with corrections to architecture; no further model instances will be created. The third condition applies to the case where the model $m_1$ has correct syntax and complies with architecture requirements, which means no additional model files are required, and the architectural review can be carried out directly to the first model off the answer codeblock.

\section{Results}

The results of the case study are presented in this section, with a focus on the modelling outputs from the generative modelling approach supported by modular context assembly. The analysis includes selected examples of model responses by LLMs, as well as a comparative analysis of model length and element counts across different treatments and models. The generative modelling approach supported by modular context assembly has been experimented on 20 parallel model runs along the workload treatments from section \ref{sec:work-treatments} and the models called in section \ref{sec:llm-treatments}. With including the corrections needed, then the modelling corpus expands to 54 - fifty-four - model instances. 

\subsection{Modelling examples}

To demonstrate the results from the case study, this section first presents selected examples of model responses by LLMs. The purpose  of characterizing these examples is to illustrate the effectiveness of the generative modelling approach, and its characteristics in proportion to the context workload. This section should present and describe the resulting models from the generative modelling approach with regard to the context workload treatments, and the LLMs called. 

\begin{figure}[htbp]
    \centering
    \rotatebox{90}{\includegraphics[width=1.3\textwidth]{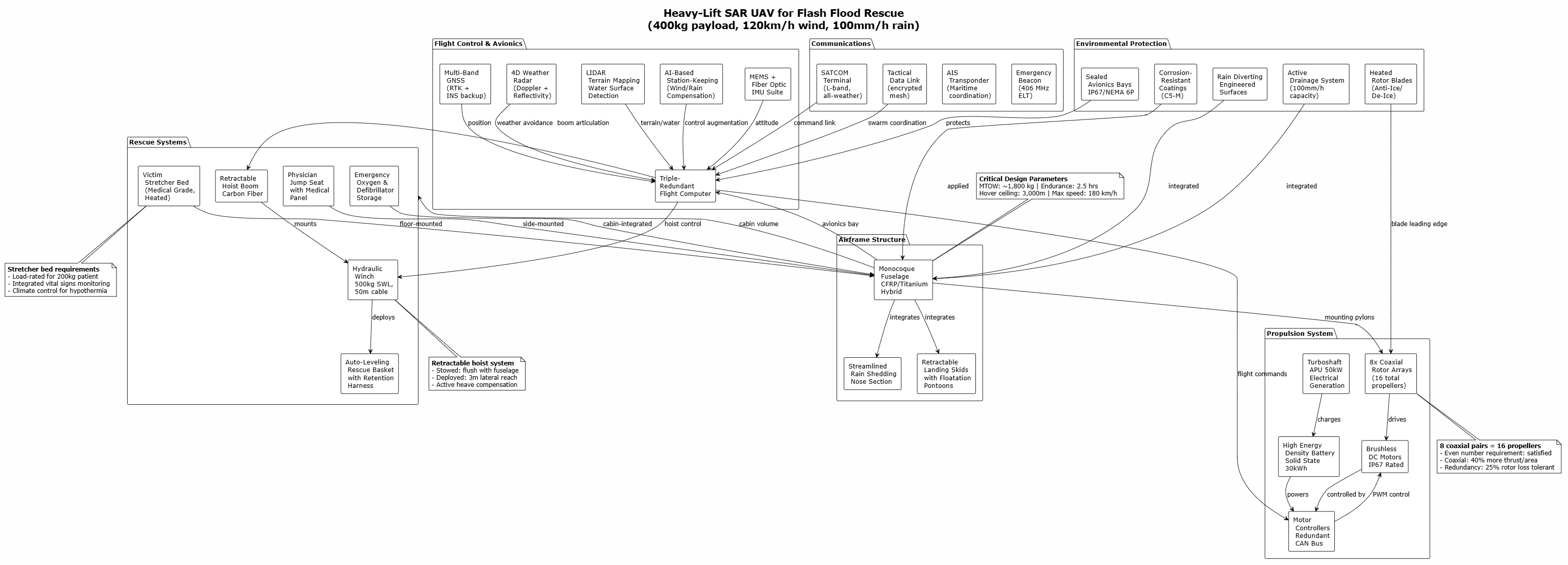}}
    \caption{Model requested from Claude Sonnet 4.6 for T1 context.}
    \label{fig:Model01}
\end{figure}

Figure \ref{fig:Model01} displays the model produced by (M5) Claude Sonnet 4.6 for the first treatment (T1) with no ancillaries to the question. This is a class model with system packages. This model has 38 high-level components with specifications in class properties (+) and 46 flows mostly expressing overall interdependence between elements. While possibly displaying a design specification of a heavy a SAR UAV, it  deviates from the intent of a component architecture. 

While the first treatment presents the question alone and therefore mostly counts on the weights from model pretraining to process the question intent, the quality of the first model is not sufficient; it does not include ports in components, neither it does specify component flows between them, then it is not a component architecture model and does not comply with the requirements in Table \ref{tab:plantuml-model-requirements}. 

\begin{figure}[htbp]
    \centering
    \includegraphics[width=0.9\textwidth]{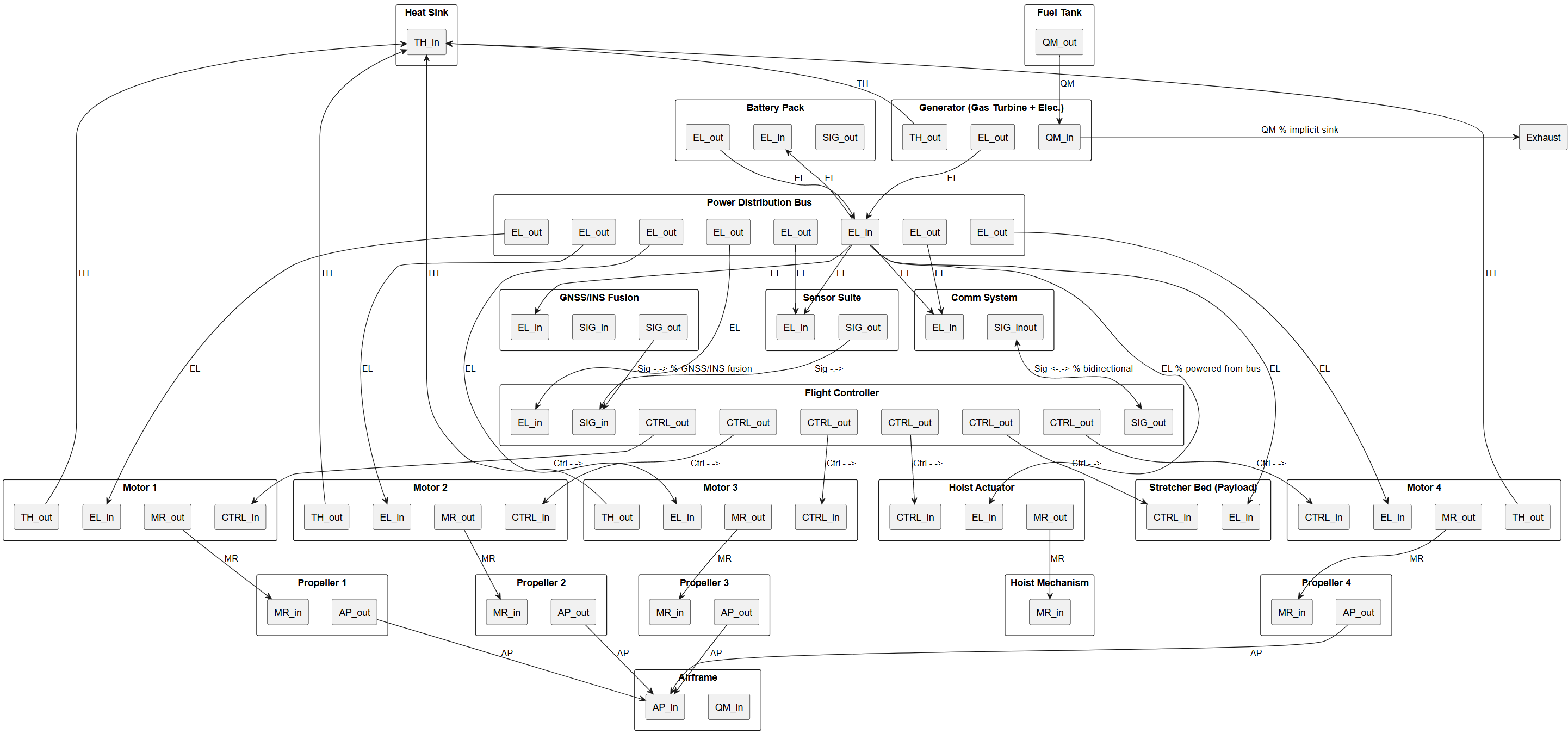}
    \caption{Model requested from Nemotron3-super for T2 context.}
    \label{fig:Model02}
\end{figure}

Then, the second treatment (T2) adds policy elements besides the question to the context workload, which is expected to improve the quality of the model output. For that purpose, the T2 treatment includes a policy statement with declaring role and directives by a $G_{sp}$ global policy component and supportive modelling guidelines by the means of a $B_{sp}$ boundary prompt component. An example model out from the T2 treatment is displayed in Figure \ref{fig:Model02}, which is a model produced by (M4) Nemotron3-super for the second treatment. 

This model has 28 individual elements and 32 flows; power transmission and distribution elements appear with two sets of soft ports, combining lower-level components. The raw model did not render correctly due to syntax errors; Figure~\ref{fig:Model02} displays the corrected \texttt{Corr1} model. Ports are represented as nested components, and flows are expressed as single declarations between port pairs. There are a few dangling ports, such as \texttt{QM\_in} within the \textit{Airframe} subsystem, and some combined components such as the generator, yet most of the model complies with the requirements in Table~\ref{tab:plantuml-model-requirements}. 

\begin{figure}[htbp]
    \centering
    \subcaptionbox{SAF Gas Engine.\label{fig:mu-template1}}{%
        \includegraphics[width=0.35\textwidth]{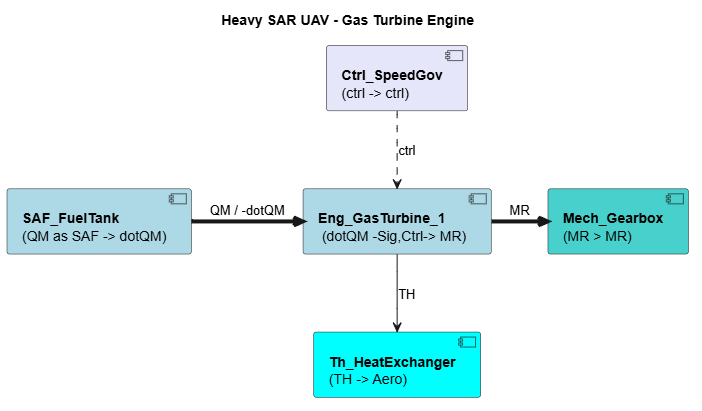}%
    }
    \hfill
    \subcaptionbox{PTO and distribution.\label{fig:mu-template2}}{%
        \includegraphics[width=0.23\textwidth]{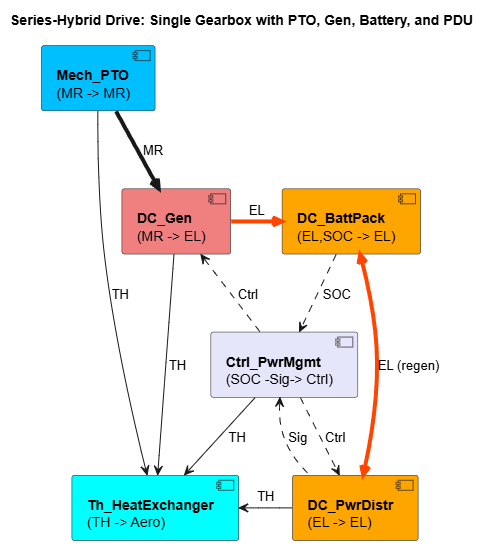}%
    }
        \hfill
    \subcaptionbox{Hoist system.\label{fig:mu-template3}}{%
        \includegraphics[width=0.21\textwidth]{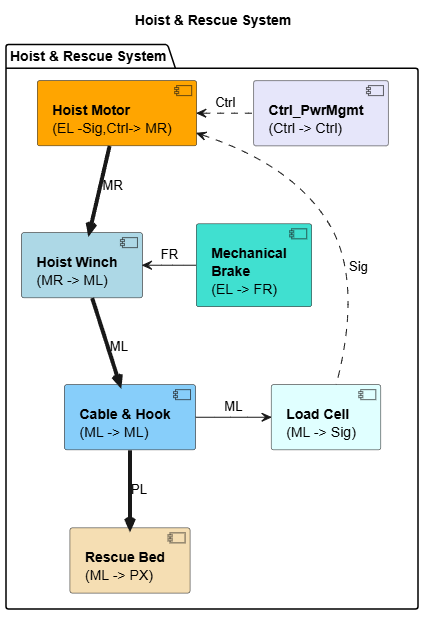}%
    }
    \caption{System model errors from $LLM_X$ answer products.}
    \label{fig:mu-templates}
\end{figure}

To support that purpose, treatments (T3) and (T4) involve the use of $\mu$-Templates \cite{Krus_LLMSAerospaceICAS_2024} as exemplars embedded in prompt vectors, with examples in figure \ref{fig:mu-template1}, \ref{fig:mu-template2} and \ref{fig:mu-template3}. These are aggregated to the query statement as prompt vectors following equation \ref{eq:eq10a} to steer the processing of the LLM. The $\mu$-Templates carry modelling-as-code snippets of the system components, ports and flows - \texttt{plantUML} in this case - to provide a reference so that the LLM shall produce a model with similar characteristics. 

\begin{figure}[htbp]
    \centering
    \includegraphics[width=0.95\textwidth]{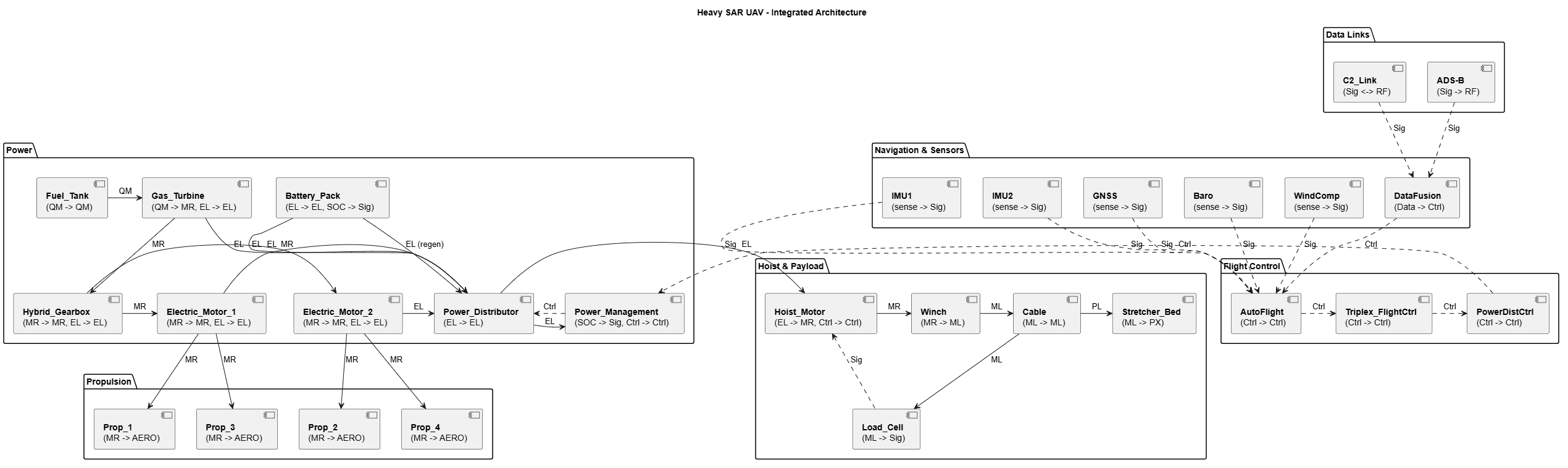}%
    \caption{Model requested from GPT-OSS-20b for T3 context.}
    \label{fig:Model03}
\end{figure}

The use of $\mu$-Templates is a key aspect of the tretments T3 and T4 in our experimental approach. To present an idea on how their use support the modelling process, Figure \ref{fig:Model03} displays the raw model produced by ChatGPT-OSS-20b for the third treatment. This T3 treatment involved the policy and prompt vector ancillaries (T3). This model has 28 individual elements and 32 flows; power transmission and distribution elements appear with two sets of soft ports -- combining lower level components. Soft port declarations are expressed within the component statements, and flows are expressed as single declarations between port pairs. 

\begin{figure}[htbp]
    \centering
    \includegraphics[width=1\textwidth]{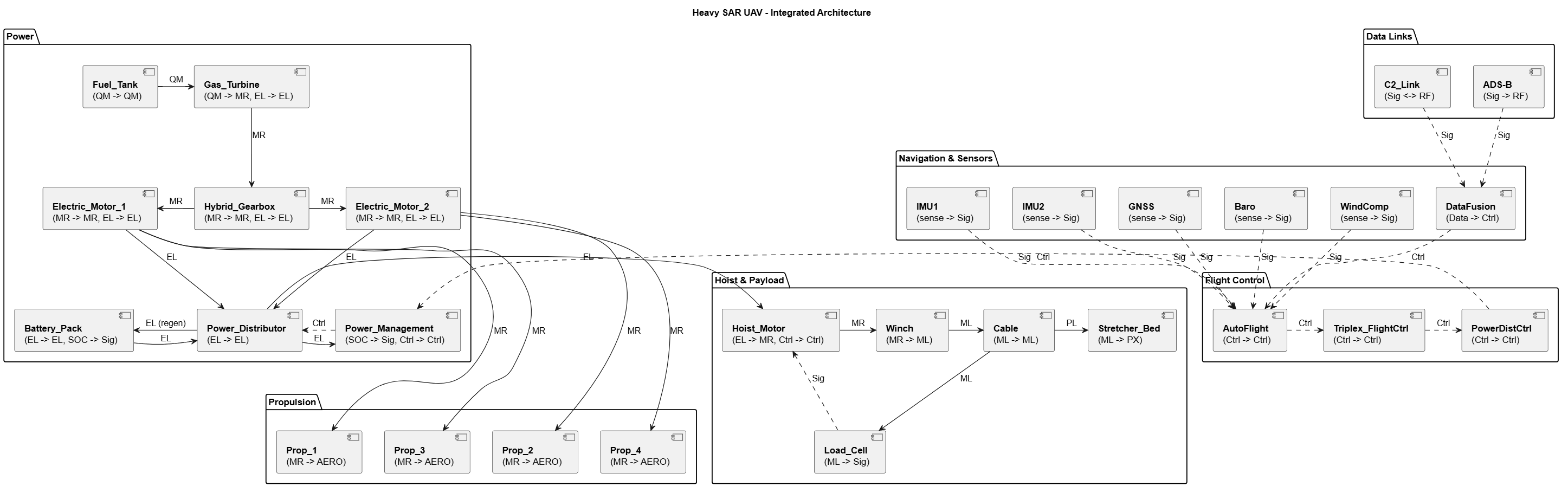}%
    \caption{Model corrected from that of GPT-OSS-20b for T3 context.}
    \label{fig:Model04}
\end{figure}

Then, Figure \ref{fig:Model04} displays a corrected model with basis on the one above. This corrective treatment involves layout modifications to the powertrain components, whereas the hoist system, electronics and communication systems remained the same, including the port-combination components, and the single flow declarations between port pairs. This means the resulting model is acceptable regarding its compliance to the requirements in Table~\ref{tab:plantuml-model-requirements}, and it is considered a valid component architecture model, worth considering for check and approval in due engineering review process.

The modelling results from the T3 treatment show that the use of $\mu$-Templates as exemplars embedded in prompt vectors can improve the quality of the model outputs from LLMs. The models produced by LLMs with the T3 treatment have more complete and accurate representations of the system components, ports, and flows, and they comply better with the requirements in Table~\ref{tab:plantuml-model-requirements}. Ultimately, the T4 treatment yields best results by a marginal difference with support of extra references; their effectiveness is somehow limited by the LLMs' ability to process and integrate the additional information, considering the increased context utilization and the effect from the bias to missing the middle.

\subsection{Modelling analysis}

This section is intended to present the results of the modelling experiment over the whole corpus and discuss its outcome with basis on the metrics in Table~\ref{tab:plantuml-model-metrics}. The analysis is carried out with respect to the context workload treatments and the LLMs called, by the means of a python tool that implements the assertions in Equations~\ref{eq:eq10}, \ref{eq:eq11} and \ref{eq:eq12}. The tool is able to parse the model files and check for syntax errors, missing elements, flows, ports, modules, end-effects, stray components, dangling ports, input/output mismatches, and chain misses. The tool also counts the number of elements, flows, ports, modules, and end-effects in the model files.

\begin{figure}[htbp]
    \centering
    \includegraphics[width=0.7\textwidth]{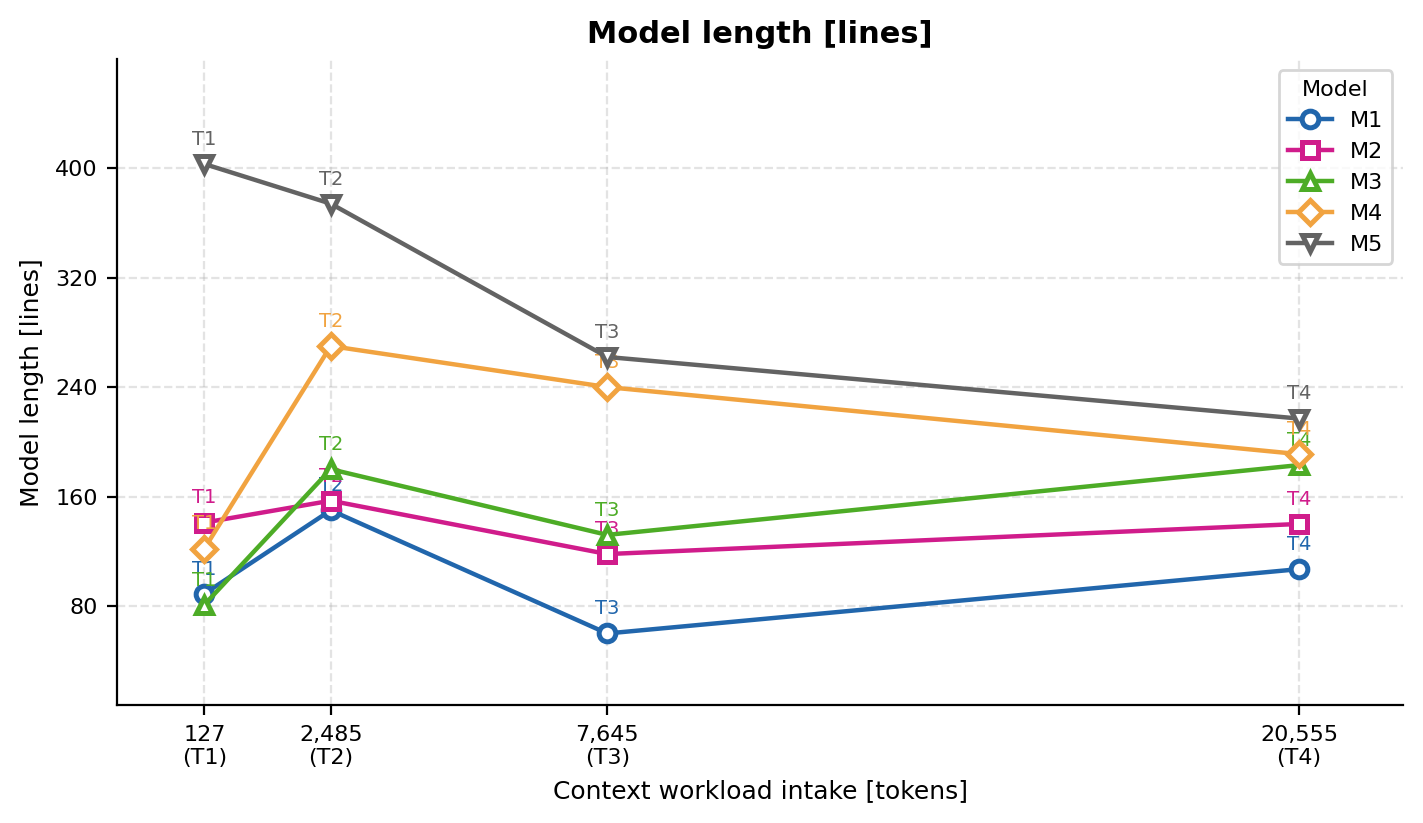}%
    \caption{Model length as result of processing context workloads.}
    \label{fig:results01}
\end{figure}

Figure \ref{fig:results01} displays the result of model length from processing context workload. Here, policy (T2) has mostly increased the modelling output from $LLM_X$ with exception of the top frontier model (M5). The introduction of $\mu$-Templates as prompt vectors into the question (T3) reduced the model length; the addition of reference (T4) did not have so significant effect, which means the added references were not so effective in steering the model output.  

\begin{figure}[htbp]
    \centering
    \subcaptionbox{Element count by workload treatment.\label{fig:metric2}}{%
        \includegraphics[width=0.49\textwidth]{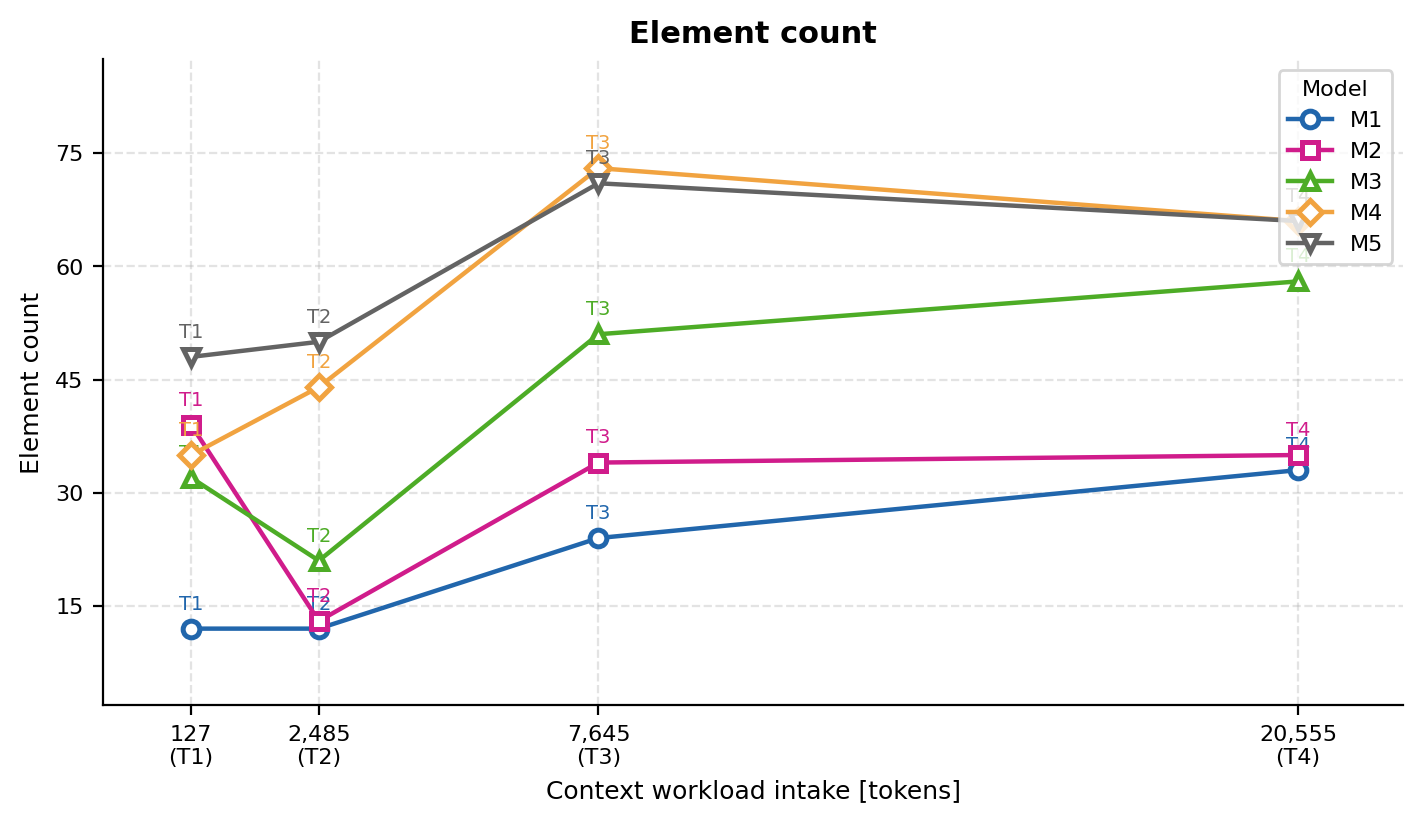}%
    }
    \hfill
    \subcaptionbox{Subelement by workload treatment.\label{fig:metric3}}{%
        \includegraphics[width=0.49\textwidth]{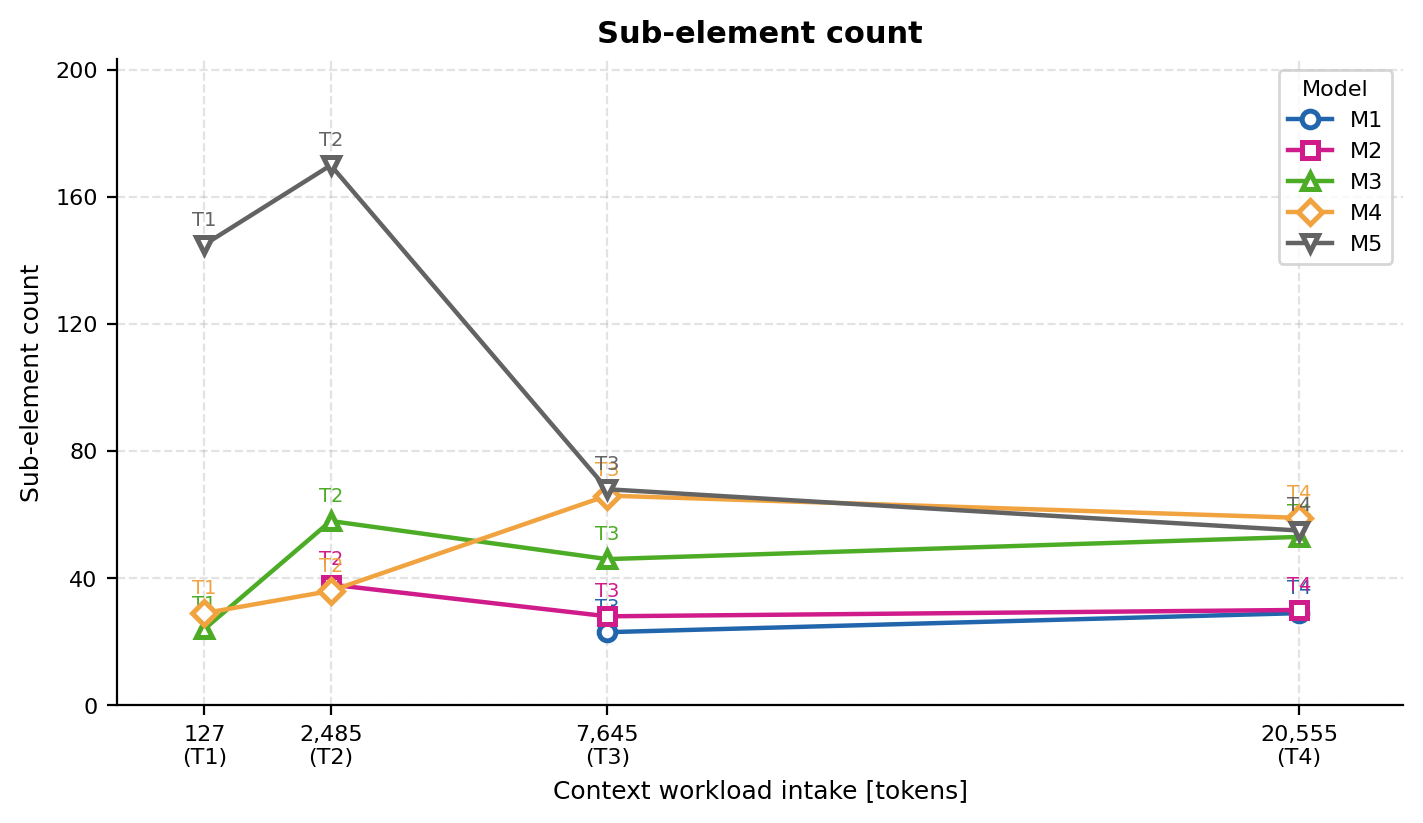}%
    }
    \caption{System model errors from $LLM_X$ answer products.}
    \label{fig:plantuml-model-validation-criteria}
\end{figure}

Then, an increase in element count from figure \ref{fig:metric2} with the introduction of policy (T2) and prompt vectors (T3) corresponds to a decrease in sub-element count from figure \ref{fig:metric3} in these treatments. Sub-elements are counted when system component statements include several component names from the vocablulary in a single entity. Then, the counts did not change significantly from treatment (T3) to treatment (T4) showing the major effect of $\mu$-Templates in steering the representation of system architecture.   

The results from the plots above are detailed in pairs regarding context workload treatments T1 \& T2 in Table \ref{tab:llm-results-t1t2} and treatments T3 \& T4 in Table \ref{tab:llm-results-t3t4}. The differences in these two sets of treatments lie in the context workload; first, that T1 and T2 treatments do not include $\mu$-Templates, while T3 and T4 treatments do; consequently, the context workload length is larger for T3 and T4 treatments as result of incorporating several $\mu$-Templates as exemplars in the prompt vectors.

Table \ref{tab:llm-results-t1t2} displays the metrics for evaluating the answer yields from $LLM_X$ on a workload that includes the single question preceded by the policy components, global and boundary prompt with different context workloads and models. Processing time differs sharply between GPU-ran and cloud-processed models, because model-processing is contrained to local GPU capacity whereas cloud-based models have access to more extensive computational resources. Then, the workload expands from 99 to 1164 tokens, thereby setting a higher load to each model's context capacity.

\begingroup
\renewcommand{\arraystretch}{1.15}
\begin{table}[htbp]
    \centering
    \caption{Metrics for evaluating $LLM_X$ answer yields on single question, treatments T1 and T2.}
    \label{tab:llm-results-t1t2}
    \scriptsize
    \resizebox{\textwidth}{!}{%
    \begin{tabular}{p{2.2cm} *{10}{c}}
        \toprule
          \multirow{2}{*}{\makecell[c]{T1 \& T2 \\ Context \\ workloads}} & \multicolumn{2}{c}{\makecell[c]{\textbf{Qwen3}\\ \textbf{8b}}} & \multicolumn{2}{c}{\makecell[c]{\textbf{ChatGPT}\\ \textbf{OSS-20b}}} & \multicolumn{2}{c}{\makecell[c]{\textbf{Nemotron3}\\ \textbf{super-120b-a12b}}} & \multicolumn{2}{c}{\makecell[c]{\textbf{Kimi}\\ \textbf{K2.5}}} & \multicolumn{2}{c}{\makecell[c]{\textbf{Claude}\\ \textbf{Sonnet4.6}}} \\
        \cmidrule(lr){2-3}
        \cmidrule(lr){4-5}
        \cmidrule(lr){6-7}
        \cmidrule(lr){8-9}
        \cmidrule(lr){10-11}
         & T1 & T2 & T1 & T2 & T1 & T2 & T1 & T2 & T1 & T2 \\
        \midrule
        \makecell[lt]{\textbf{Answer} \textnormal{[Tk]}} & 1207 & 2038 & 1039 & 1718 & 1241 & 1790 & 1600 & 3080 & 4039 & 5185 \\
        \addlinespace[3pt]
        \makecell[lt]{\textbf{Utilization} \textnormal{[\%]}} & 4,07 & 13,81 & 3,56 & 12,83 & 2,09 & 6,52 & 0,67 & 2,17 & 0,42 & 0,77 \\
        \addlinespace[3pt]
        \makecell[lt]{\textbf{Time} \textnormal{[min: sec]}} & 6:19 & 12:42 & 0:12 & 0:41 & 1:31 & 2:34 & 0:14 & 1:01 & 1:37 & 1:48 \\
        \addlinespace[3pt]
        \makecell[lt]{\textbf{Model type/}\\ \textbf{Length}\space[lines]} & \makecell[lt]{Cmp\\89} & \makecell[lt]{Cmp\\150} & \makecell[lt]{\textsuperscript{*}Cmp\\141} & \makecell[lt]{Cmp\\157} & \makecell[lt]{\textsuperscript{*}Cmp\\81} & \makecell[lt]{Cmp\\180} & \makecell[lt]{\textsuperscript{*}Cmp\\122} &\makecell[lt]{\textsuperscript{*}Cmp\\270} & \makecell[lt]{Cls\\403} & \makecell[lt]{\textsuperscript{*}Cmp\\374} \\
        \addlinespace[3pt]
        \makecell[lt]{\textbf{Element type/}\\ \textbf{count} \textnormal{[:n]}} & \makecell[lt]{Sub\\12} & \makecell[lt]{Sub\\12} & \makecell[lt]{Unit\\39} & \makecell[lt]{Sub\\13} & \makecell[lt]{Sub\\32} & \makecell[lt]{Unit\\21} & \makecell[lt]{Sub\\35} & \makecell[lt]{Sub\\44} & \makecell[lt]{Sub\\48} & \makecell[lt]{Unit\\50} \\
        \addlinespace[3pt]
        \makecell[lt]{\textbf{Flow type/}\\ \textbf{count} \textnormal{[:n]}} & \makecell[lt]{Fu-Ct\\22} & \makecell[lt]{n/a\\43} & \makecell[lt]{n/a\\32} & \makecell[lt]{Fw-Fu\\32} & \makecell[lt]{Fu-Ct\\24} & \makecell[lt]{Fw-Fu\\40} & \makecell[lt]{Fu-Ct\\30} & \makecell[lt]{Fw-Fu\\41} & \makecell[lt]{Fu-Ct\\46} & \makecell[lt]{Fw-Fu\\101} \\
        \addlinespace[3pt]
        \makecell[lt]{\textbf{Subelements/}\\ \textbf{count} \textnormal{[:n]}} & \makecell[lt]{-} & \makecell[lt]{-} & \makecell[lt]{-} & \makecell[lt]{Port\\38} & \makecell[lt]{Spec\\24} & \makecell[lt]{Port\\58} & \makecell[lt]{Spec\\29} & \makecell[lt]{Spec\\36} & \makecell[lt]{Spec\\145} & \makecell[lt]{Port*\\170} \\
        \addlinespace[3pt]
        \makecell[lt]{\textbf{Elements/}\\ \textbf{Ports/}\\ \textbf{Flows} \textnormal{[:n]}} & \makecell[lt]{12\\0\\22} & \makecell[lt]{12\\0\\43} & \makecell[lt]{39\\0\\32} & \makecell[lt]{13\\38\\32} & \makecell[lt]{32\\0\\24} & \makecell[lt]{21\\58\\40} & \makecell[lt]{35\\0\\30} & \makecell[lt]{44\\0\\41} & \makecell[lt]{48\\0\\46} & \makecell[lt]{50\\170\\101} \\
        \addlinespace[3pt]
        \makecell[lt]{\textbf{Syntax}\\ \textbf{misses}\\ \textnormal{[Types/:n]}} & none & none & none & Fl: 16 & none & \makecell[lt]{Fl: 40\\Pt: 2} & none & Fl: 30 & none & none \\
        \addlinespace[3pt]
        \makecell[lt]{\textbf{Element}\\ \textbf{misses}\\ \textnormal{[Types/:n]}} & 23 & 22 & 14 & 13 & 10 & 16 & 12 & 15 & 7 & 12 \\
        \addlinespace[3pt]
        \makecell[lt]{\textbf{Port/flow}\\ \textbf{mismatches}\\ \textnormal{[Types/:n]}} & Ph: 1 & none & Orph: 3 & Dang: 38 & Orph: 1 & Dir: 7 & \makecell[lt]{Orph: 2\\Ph: 1} & Orph: 2 & Orph: 1 & Dang: 9 \\
        \bottomrule
    \end{tabular}%
    }
    \\[2pt]
    {\scriptsize\justifying
    \textit{Legend.} \\ \textbf{Model type:} Cmp~--~component diagram; \textsuperscript{*}Cmp~--~packaged components (components grouped across multiple subsystem packages); Cls~--~class diagram. \\
    \textbf{Element type:} Sub~--~subsystem element; Unit~--~unit element; Spec~--~named property/rating; Port~--~connection in code; Port\textsuperscript{*}~--~connection in element.\\
    \textbf{Flow type:} Fu~--~function; Ct~--~containment; Dv~--~deliverable; Fw~--~port-matching flow; Fu-Ct~--~mixed function/containment; n/a~--~unlabeled flows.\\
    \textbf{Counts:} length in code lines; Element/Subelement/Flow counts as [:n].\\
    \textit{Audit.} \\
    \textbf{Syntax misses} are render-blocking errors (those requiring a syntax-only \textit{Corr1} pass): Fl~--~flow statement, Pt~--~port statement, El~--~declaration. \\
    \textbf{Element misses} counted strictly vs the SAR reference checklist (per fine-grained element, of~32). \\
    \textbf{Mismatches}: Orph~--~declared element, no flow; Dang~--~declared, unwired; Dir~--~direction/port error; Ph~--~flow to/from undeclared id. 
     \par}
\end{table}
\endgroup

Answer size varies substantially, from 466 tokens (Qwen3 8B) to 1,984 tokens (Claude Sonnet 4.6), indicating different expansion behavior even under comparable prompting conditions. Model-output structure becomes richer in larger responses, as seen in line counts (31 to 195) and flow counts (12 to 56), suggesting greater elaboration of system interactions. Processing time also differs sharply, with local/smaller models taking longer in this setup (up to 12:42) while larger cloud models respond faster (down to 0:14). 

Besides computing hardware itself, the cloud processing time is determined by the scale of processing resource and the bandwidth of the cloud connection. Regarding the answer output, model-output structure becomes richer in larger responses, as seen in line counts (31 to 195) and flow counts (12 to 56), suggesting greater elaboration of system interactions. 

Overall, the results suggest that higher-capacity models improve output depth and structural detail, while all models remain lightly loaded in context usage for these treatments. This means that the answers depend more of the model pretraining than of the context workload, and thereby the model outputs are more diverse in structure and content. 

Table \ref{tab:llm-results-t3t4} displays the metrics for evaluating the answer yields from $LLM_X$ on a workload that includes the single question preceded by the policy components, global and boundary prompt with different context workloads and models; $\mu$-templates are included in prompt vectors along the context workload in both T3 and T4, whereas T4 also includes additional context from references to previous flash-flood events and SAR resources involved. 

The workload expands to 7645 tokens and then to 20555 tokens, thereby setting a significantly higher utilization of each model's context capacity by the incoming context workload.

\begingroup
\renewcommand{\arraystretch}{1.15}
\begin{table}[htbp]
    \centering
    \caption{Metrics for evaluating $LLM_X$ answer yields on single question, treatments T3 and T4.}
    \label{tab:llm-results-t3t4}
    \scriptsize
    \resizebox{\textwidth}{!}{%
    \begin{tabular}{p{2.2cm} *{10}{c}}
        \toprule
          \multirow{2}{*}{\makecell[c]{T3 \& T4 \\ Context \\ workloads}} & \multicolumn{2}{c}{\makecell[c]{\textbf{Qwen3}\\ \textbf{8b}}} & \multicolumn{2}{c}{\makecell[c]{\textbf{ChatGPT}\\ \textbf{OSS-20b}}} & \multicolumn{2}{c}{\makecell[c]{\textbf{Nemotron3}\\ \textbf{super-120b-a12b}}} & \multicolumn{2}{c}{\makecell[c]{\textbf{Kimi}\\ \textbf{K2.5}}} & \multicolumn{2}{c}{\makecell[c]{\textbf{Claude}\\ \textbf{Sonnet4.6}}} \\
        \cmidrule(lr){2-3}
        \cmidrule(lr){4-5}
        \cmidrule(lr){6-7}
        \cmidrule(lr){8-9}
        \cmidrule(lr){10-11}
         & T3 & T4 & T3 & T4 & T3 & T4 & T3 & T4 & T3 & T4 \\
        \midrule
        \makecell[lt]{\textbf{Answer} \textnormal{[Tk]}} & 1285 & 1245 & 1808 & 2095 & 1997 & 2425 & 3189 & 2964 & 3897 & 3379 \\
        \addlinespace[3pt]
        \makecell[lt]{\textbf{Utilization} \textnormal{[\%]}} & 27,25 & 66,53 & 28,85 & 69,12 & 14,71 & 35,06 & 4,23 & 9,19 & 1,15 & 2,39 \\
        \addlinespace[3pt]
        \makecell[lt]{\textbf{Time} \textnormal{[min: sec]}} & 8:57 & 8:42 & 0:15 & 0:26 & 0:20 & 2:49 & 0:25 & 0:23 & 1:39 & 1:30 \\
        \addlinespace[3pt]
        \makecell[lt]{\textbf{Model type/}\\ \textbf{Length}\space[lines]} & \makecell[lt]{Cmp\\60} & \makecell[lt]{\textsuperscript{*}Cmp\\107} & \makecell[lt]{\textsuperscript{*}Cmp\\118} & \makecell[lt]{\textsuperscript{*}Cmp\\140} & \makecell[lt]{\textsuperscript{*}Cmp\\132} & \makecell[lt]{\textsuperscript{*}Cmp\\183} & \makecell[lt]{\textsuperscript{*}Cmp\\240} & \makecell[lt]{\textsuperscript{*}Cmp\\191} & \makecell[lt]{Cmp\\262} & \makecell[lt]{\textsuperscript{*}Cmp\\217} \\
        \addlinespace[3pt]
        \makecell[lt]{\textbf{Element type/}\\ \textbf{count} \textnormal{[:n]}} & \makecell[lt]{Unit\\24} & \makecell[lt]{Unit\\33} & \makecell[lt]{Unit\\34} & \makecell[lt]{Unit\\35} & \makecell[lt]{Unit\\51} & \makecell[lt]{Unit\\58} & \makecell[lt]{Unit\\73} & \makecell[lt]{Unit\\66} & \makecell[lt]{Unit\\71} & \makecell[lt]{Unit\\66} \\
        \addlinespace[3pt]
        \makecell[lt]{\textbf{Flow type/}\\ \textbf{count} \textnormal{[:n]}} & \makecell[lt]{Fw-Fu\\21} & \makecell[lt]{Fw-Fu\\36} & \makecell[lt]{Fw-Fu\\32} & \makecell[lt]{Fw-Fu\\49} & \makecell[lt]{Fw-Fu\\70} & \makecell[lt]{Fw-Fu\\70} & \makecell[lt]{Fw-Fu\\109} & \makecell[lt]{Fw-Fu\\82} & \makecell[lt]{Fw-Fu\\108} & \makecell[lt]{Fw-Fu\\74} \\
        \addlinespace[3pt]
        \makecell[lt]{\textbf{Subelements/}\\ \textbf{count} \textnormal{[:n]}} & \makecell[lt]{Port\\23} & \makecell[lt]{Port\\29} & \makecell[lt]{Port\\28} & \makecell[lt]{Port\\30} & \makecell[lt]{Port\\46} & \makecell[lt]{Port\\53} & \makecell[lt]{Port\\66} & \makecell[lt]{Port\\59} & \makecell[lt]{Port\\68} & \makecell[lt]{Port\\55} \\
        \addlinespace[3pt]
        \makecell[lt]{\textbf{Syntax misses}\\ \textnormal{[Types/:n]}} & none & none & Fl: 8 & Pre: 1 & none & none & none & Fl: 2 & none & none \\
        \addlinespace[3pt]
        \makecell[lt]{\textbf{Element misses}\\ \textnormal{[Types/:n]}} & 16 & 13 & 13 & 9 & 2 & none & none & none & none & none \\
        \addlinespace[3pt]
        \makecell[lt]{\textbf{Mismatches}\\ \textnormal{[Types/:n]}} & none & none & none & none & \makecell[lt]{Ph: 3\\Dup: 8} & none & none & none & none & none \\
        \bottomrule
    \end{tabular}%
    }
    \\[2pt]
    {\scriptsize\justifying
    \textit{Legend.} \\ \textbf{Model type:} Cmp~--~component diagram; \textsuperscript{*}Cmp~--~packaged components (components grouped across multiple subsystem packages); Cls~--~class diagram. \\
    \textbf{Element type:} Sub~--~subsystem element; Unit~--~unit element; Spec~--~named property/rating; Port~--~connection in code; Port\textsuperscript{*}~--~connection in element.\\
    \textbf{Flow type:} Fu~--~function; Ct~--~containment; Dv~--~deliverable; Fw~--~port-matching flow; Fu-Ct~--~mixed function/containment; n/a~--~unlabeled flows.\\
    \textbf{Counts:} length in code lines; Element/Subelement/Flow counts as [:n].\\
    \textit{Audit.} \\
    \textbf{Syntax misses} are render-blocking errors (those requiring a syntax-only \textit{Corr1} pass): Fl~--~flow statement, Pt~--~port statement, El~--~declaration. \\
    \textbf{Element misses} counted strictly vs the SAR reference checklist (per fine-grained element, of~32). \\
    \textbf{Mismatches}: Orph~--~declared element, no flow; Dang~--~declared, unwired; Dir~--~direction/port error; Ph~--~flow to/from undeclared id. 
     \par}
\end{table}
\endgroup

Processing time also differs sharply, with local/smaller models taking longer in this setup (up to 08:21) while larger cloud models respond faster (down to 0:30). The trends between local and cloud models are consistent with the overall performance patterns observed, yet the presence of examples and references seemed to enable better performance to timing of local models.

Answer size is somehow correlated with the parameter complexity of the models in use, with Qwen3-8b yielding ~1300 tokens and the frontier models, Kimi and Claude yielding ~3000 tokens.Model-output structure becomes richer in larger responses, as seen in line counts (31 to 195) and flow counts (21 to 109), suggesting greater elaboration of system interactions and interconnections in larger models - the interdependencies between component functions may appear more pronounced from larger models. 

Overall, the results suggest that higher-capacity models improve output depth and structural detail, while all models remain lightly loaded in context usage for this experiment. At the same time, fromtier models manage to yield more complete and accurate representations of the system components, ports, and flows, and the results from the Table show that they perform better at leveraging context workload onto complying with the requirements in Table~\ref{tab:plantuml-model-requirements}.

The most significant gain in model quality is observed when $\mu$-Templates are used in conjunction with policy and prompt vector ancillaries (T3), as compared to the baseline treatments (T1 and T2). Then, the \textit{lost-in-the-middle} effect limits the effectiveness of the additional references in T4, as the model's attention will concentrate on the policy by the beginning of the workload and on the $\mu$-Templates along the core question by the end of the workload.

There are significant effects from treatments regarding element misses and port/flow mismatches from $LLM_X$ yields. Element misses were consistently high on smaller LLMs, and policy and guidelines along the question (T2) made component misses to increase. Then, $\mu$-Templates as prompt vectors practically annulled component misses in (T3) and (T4) treatments.   

\subsection{Discussion of results}

This paper presented modelling samples towards architecting a heavy SAR UAV system, a set of requirements and associated assertions on which the models are evaluated, and results from a quantitative analysis over the generated models. This case study experiment involved our running 20 queries spanning context workloads and selected LLMs that were requested to generate system architecture \verb+plantUML+ codeblock models. 

The overall modelling answer corpus evolved from 20 treatment answer results to 55 individual model files, by counting the answer files plus the spawned models from the answer codeblocks upon the corrections required by the eval assertions. The analysis part involved setting up modelling metrics to verify the output of the LLMs; here, we focus element count and syntax requirements for the verification.  

\begin{figure}[htbp]
    \centering
    \includegraphics[width=0.7\textwidth]{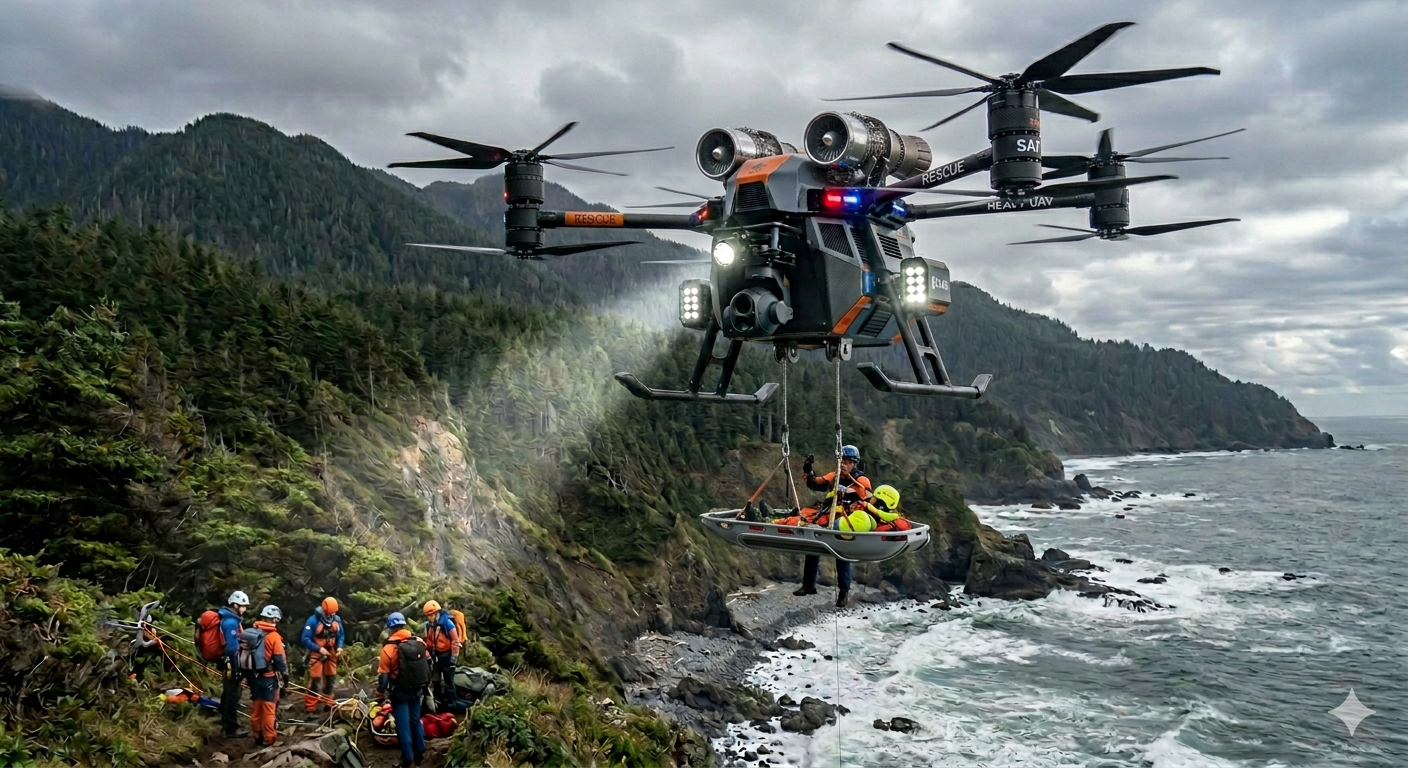}%
    \caption{AI-generated Heavy SAR UAV rendering (source: Gemini).}
    \label{fig:gemini-render}
\end{figure}

The whole dataset reveals the effect of the primacy and recency, characteristics of information units within a context workload to an LLM that were revealed by \cite{liu_lostinmiddle_2017} upon finding out on the limitations of LLMs under high utilization of their context length capacity. Here, smaller models such as Qwen3-8b (M1) and GPT-OSS-20b (M2), both with maximum 32767 Tk of context capacity, were significantly utilized -- especially in treatments T3 and T4 with including $\mu$-Template exemplars.

These effects have something to do with context units that \emph{precede} the question having a divergence effect, and context units that are within the question, which actually \emph{succeed} the query statement -- that carries the actual request and thereby the modelling intent -- and make both convergence and enforcement effects to expected patterns in both coding and modelling. This means that the effects of \emph{primacy} and \emph{recency} as discussed by \cite{liu_lostinmiddle_2017} manifest themselves when the context workload in single question becomes significant towards context length capacity. 

While the performance of frontier models such as Kimi K2.5 (M4) and Claude Sonnet 4.6 (M5) in modelling architectures with support of $\mu$-Templates -- and that of similar-scale models such as Deepseek and ChatGPT-5.x from previous experience of ours in \cite{mariniKrus_PromptComposition_2024} -- bears no remarks upon their higher scale and longer context capacities, $\mu$-templates have a convergence effect to a desired model setup regarding its format, structure and compliance. In models such as Qwen3, GPT-OSS and Nemotron, the effects of context operations deserve attention. 

Their performance comes not without their responses coming briefly, either with subcomponents declared within component blocks -- which deviate from the purpose of architecting -- or with omitting components entirely yet complying with code syntax and modelling conventions. Here, one expects a topology of components, effect flows and ports that supports component-level specification and designing with mind to later integration tasks. At the same time, such responses can be fixed to completeness, accuracy and compliance with a few minutes of juggling model code.

One significant matter of attention to be considered is the effect of different units onto the workload being forwarded to the LLMs. The workload units are the policy, the prompt vectors, the references and the question itself. The policy units are intended to set up a context for the modelling request, and they are expected to be processed by the LLMs as a \emph{primacy} effect. The prompt vectors are intended to set up a context for the modelling request, and they are expected to be processed by the LLMs as a \emph{recency} effect. 

Here, references in the \textbf{\textit{**middle**}} of the context workload are expected to be processed by the LLMs as a \emph{convergence} effect, yet their effectiveness is limited by the \emph{lost-in-the-middle} effect. Then, this means that policy and modelling directives, as well as examples, are the key elements on which the LLMs focus their attention to produce a modelling result. Nevertheless, in-context learning could be a powerful technique regarding the build-up of reference for modelling requests, whose usefulness goes down to sorting out the token economics over searching through a knowledge database.

The use of evaluation assertions to verify the modelling results from LLMs is a key element onto enabling the usefulness of a modelling corpus onto the development of a learning database for systems modelling with generative artificial intelligence. Along with context handling, evaluations enable setting convergence criteria for modelling loops and ultimately enable the potential for the use of hybrid deterministic-probabilistic modelling approaches to systems design, such as proposed by \cite{krus_augmenting_2025} when requesting a deterministic model builder algorithm from a context prompt.

Understanding the capabilities of LLMs of a scale spectrum from \emph{giga} to \emph{tera}-scale enables proceeding to investigate and take advantage of novel context handling techniques to improve the capability of artificial intelligence onto systems modelling to design intent.

\section{Conclusions}
This contribution has presented a framework of context operations on which to prompt at large language models. Being implemented in a chatbox applicaiton as that shown by \cite{mariniKrus_PromptComposition_2024}, this framework was successfully operated towards a case study of aircraft design, here being a heavy SAR UAV. This case study contributes with evolving from modelling examples towards the experimentation with modelling requests, and the verification of their modelling results through measuring their architectural properties. 

The evaluation of modelling quality by large language models from the synthesis of modular context workloads enabled us to discuss on the modelling capability of LLMs under a single query. The modelling results presented from the experiment are on par with current understanding about the workings of LLMs to system design. At the same time, these modelling results enable the development of a learning database on how to support systems modelling with generative artificial intelligence. 

Future work involves expanding the modular workload approach to conversation threading and agentic modelling, which in turn requires the expansion of the learning base to fine-tune model counting and validation rules. 

\section{Acknowledgments}
This study has been carried out with the support of the CNPq-CISB grant no. 200944/2024-0 of the Brazilian National Council for Scientific and Technological Development (CNPq) and the Swedish-Brazilian Centre for Innovation, through collaborative research work carried out by the authors at the Federal University of Santa Maria (UFSM) in Brazil and at the Linköping University (LiU) in Sweden.

The authors manifest their gratitude in advance to reviewers for their valuable comments and suggestions, which help improve the quality of this paper.

\bibliographystyle{unsrtnat}
\bibliography{references}

@inproceedings{cummings_AutomationBias_2004,
    author          = {Mary L. Cummings},
    title           = {Automation Bias in Intelligent Time Critical Decision Support Systems},
    booktitle       = {: 1st AIAA\\ Intelligent Systems Technical Conference},
    chapter         = {},
    pages           = {},
    year            = {2004},
    organization    = {Chicago, IL: American Institute of Aeronautics and Astronautics},
    doi             = {10.2514/6.2004-6313}
}

@article{torngren_Complexity_2018,
    title           = {How to deal with the complexity of future cyber-physical systems?},
    author          = {T{\"o}rngren, Martin and Grogan, Paul T.},
    journal         = {Designs},
    issn            = {2411-9660},
    volume          = {2},
    number          = {4},
    pages           = {40 pp.},
    year            = {2018},
    publisher       = {MDPI},
    doi             = {10.3390/designs2040040}
}

@article{grogan_perception_2021,
	title          = {Perception of {Complexity} in {Engineering} {Design}},
	volume         = {24},
	issn           = {1520-6858},
	doi            = {10.1002/sys.21574},
	language       = {en},
	number         = {4},
	journal        = {Systems Engineering},
	author         = {Grogan, Paul T},
	year           = {2021},
	pages          = {221--233}
}

@article{brown_language_2020,
    title             = {Language models are few-shot learners},
    author            = {Brown, Tom and Mann, Benjamin and Ryder, Nick and Subbiah, Melanie and Kaplan, Jared D and Dhariwal, Prafulla and Neelakantan, Arvind and Shyam, Pranav and Sastry, Girish and Askell, Amanda and others},
    journal           = {Advances in neural information processing systems},
    volume            = {33},
    pages             = {1877--1901},
    year              = {2020},
    url               = {https://proceedings.neurips.cc/paper_files/paper/2020/hash/1457c0d6bfcb4967418bfb8ac142f64a-Abstract.html}
}

@article{vaswani_attention_2017,
    title             = {Attention is all you need},
    author            = {Vaswani, Ashish and Shazeer, Noam and Parmar, Niki and Uszkoreit, Jakob and Jones, Llion and Gomez, Aidan N and Kaiser, {\L}ukasz and Polosukhin, Illia},
    journal           = {Advances in neural information processing systems},
    volume            = {30},
    year              = {2017},
    url               = {https://proceedings.neurips.cc/paper_files/paper/2017/hash/3f5ee243547dee91fbd053c1c4a845aa-Abstract.html}
}

@inproceedings{Krus_LLMSAerospaceICAS_2024,
    author          = {Krus, Petter},
    booktitle       = {: 34th Congress of the International Council\\ of the Aeronautical Sciences, ICAS 2024},
    title           = {{Large language model in aircraft system design}},
    year            = {2024},
    url             = {https://www.icas.org/icas_archive/icas2024/data/papers/icas2024_0514_paper.pdf}
}

@article{johnsetal_LLMstoMBSE_2024,
    author          = {Johns, Brian and Carroll, Kristina and Medina, Casey 
                       and others},
    title           = {{AI} Systems Modeling Enhancer {(AI-SME)}: Initial Investigations into a ChatGPT-enabled {MBSE} Modeling Assistant},
    journal         = {{INCOSE} International Symposium},
    issn            = {2334-5837},
    volume          = {34},
    number          = {1},
    pages           = {1149-1168},
    doi             = {10.1002/iis2.13201},
    year            = {2024}
}

@article{teubner_welcome_2023,
    title             = {Welcome to the Era of {C}hat{GPT} et al.},
    author            = {Teubner, Timm and Flath, Christoph M and Weinhardt, Christof and Van Der Aalst, Wil and Hinz, Oliver},
    journal           = {Business \& information systems engineering},
    volume            = {65},
    number            = {2},
    pages             = {95--101},
    year              = {2023},
    issn              = {1867-0202},
    doi               = {10.1007/s12599-023-00795-x},
    publisher         = {Springer Fachmedien Wiesbaden}
}

@techreport{graydon_UsesofLLMs_2025,
    type            = {Technical Memorandum},
    key             = {2025-0001849},
    title           = {Examining Proposed Uses of {LLMs} to Produce or Assess Assurance Arguments},
    author          = {Graydon, Mallory S. and Lehman, Sarah M.},
    institution     = {NASA/TM-2025-0001849, National Aeronautics and Space Administration},
    month           = mar,
    year            = {2025},
    url             = {https://ntrs.nasa.gov/api/citations/20250001849/downloads/NASA-TM-20250001849.pdf}
}

@inproceedings{gomez_LLMs_2024,
    title           = {Large language models in complex system design},
    author          = {Pradas-Gomez, Alejandro and Krus, Petter 
                       and Panarotto, Massimo and Isaksson, Ola},
    booktitle       = {: {DESIGN 2024} {I}nternational design conference}, 
    volume          = {4},
    pages           = {pp. 2197--2206},
    year            = {2024},
    organization    = {Cavtat, Croatia: Design Society},
    doi             = {10.1017/pds.2024.222}
}

@inproceedings{marinietal_Human-machine_2025,
    title           = {Context of collaborative human-machine systems architecture design for enhanced functionality awareness and balanced command and control authority},
    author          = {Marini, Vinicius K. and Alfredson, Jens and Krus, Petter},
    booktitle       = {Proceedings of the 12th Swedish Aersopace Technology Congress - FT2025},
    organization    = {Stockholm, Sweden: Swedish Society for Aeronautics and Astronautics (FTF)},
    year            = {2025},
    doi             = {10.3384/ecp215.1192} 
}

@article{camara_Assessment_2023,
    title           = {On the assessment of generative {AI} in modeling tasks: an experience report with {ChatGPT} and {UML}},
    author          = {Cámara, Javier and Troya, Javier and Burgueño, Lola and Vallecillo, Antonio},
    volume          = {22},
    issn            = {1619-1374},
    doi             = {10.1007/s10270-023-01105-5},
    number          = {3},
    journal         = {Software and Systems Modeling},
    month           = jun,
    year            = {2023},
    pages           = {781--793},
}

@article{dehart_LLM&SysML_2024,
    author          = {DeHart, John K.},
    title           = {Leveraging Large Language Models for Direct Interaction with {SysML v2}},
    journal         = {INCOSE International Symposium},
    issn            = {2334-5837},
    volume          = {34},
    number          = {1},
    pages           = {2168-2185},
    doi             = {10.1002/iis2.13262},
    year            = {2024}
}

@article{timperley_assessment_2025,
	title          = {Assessment of large language models for use in generative design of model based spacecraft system architectures},
	volume         = {36},
	issn           = {0954-4828},
	number         = {4},
	journal        = {Journal of Engineering Design},
	publisher      = {Taylor \& Francis},
	author         = {Timperley, Louis Richard and Berthoud, Lucy and Snider, Chris and Tryfonas, Theo},
	year           = {2025},
    doi            = {10.1080/09544828.2025.2453401},
	pages          = {550--570},
}

@article{baluetal_LLM-RAG_2025, 
    title           = {Towards Automated Safety Requirements Derivation Using Agent-based {RAG}}, 
    volume          = {5}, 
    DOI             = {10.1609/aaaiss.v5i1.35605}, 
    number          = {1}, 
    journal         = {Proceedings of the AAAI Symposium Series}, 
    author          = {Balu, Balahari Vignesh and Geissler, Florian and Carella, Francesco 
                      and others}, 
    year            = {2025}, 
    month           = {May}, 
    pages           = {299-307} 
}

@article{hanke_AIAugmentedSE_2025,
    title           = {\\{AI}-augmented systems engineering: conceptual application of retrieval-augmented generation for model-based systems engineering graph},
    author          = {Hanke, Fabian and Bita, Isaac Mpidi and von Hei{\ss}en, Oliver and Julian, Weller and Aschot, Hovemann and Roman, Dumitrescu},
    journal         = {Proceedings of the Design Society},
    volume          = {5},
    pages           = {439--448},
    year            = {2025},
    doi             = {10.1017/pds.2025.10058},
    publisher       = {Cambridge University Press}
}

@inproceedings{nouri_LLM-Req_2024,
    title             = {Engineering safety requirements for autonomous driving with large language models},
    author            = {Nouri, Ali and Cabrero-Daniel, Beatriz and T{\"o}rner, Fredrik 
                    and others},
    booktitle         = {2024 IEEE 32nd International Requirements Engineering Conference (RE)},
    pages             = {218--228},
    year              = {2024},
    doi               = {10.1109/RE59067.2024.00029},
    organization      = {IEEE}
}

@article{elHassani_integratingLLM-FMEA_2024,
    title           = {Integrating large language models for improved failure mode and effects analysis (FMEA): a framework and case study},
    author          = {El Hassani, Ibtissam and Masrour, Tawfik and Kourouma, Nouhan and Motte, Damien and Tav{\v{c}}ar, Jo{\v{z}}e},
    journal         = {Proceedings of the Design Society},
    volume          = {4},
    pages           = {2019--2028},
    year            = {2024},
    doi             = {10.1017/pds.2024.204},
    publisher       = {Cambridge University Press}
}

@article{elHassani_integratingLLM-FMEA_2025,
    title           = {{AI}-driven {FMEA}: integration of large language models for faster and more accurate risk analysis},
    author          = {El Hassani, Ibtissam and Masrour, Tawfik and Kourouma, Nouhan and Tav{\v{c}}ar, Jo{\v{z}}e},
    journal         = {Design Science},
    volume          = {11},
    issn            = {2053-4701},
    pages           = {e10},
    year            = {2025},
    doi             = {10.1017/dsj.2025.7},
    publisher       = {Cambridge University Press}
}

@article{qi_STPA-GPT_2025,
    title           = {Safety analysis in the era of large language models: A case study of {STPA} using {ChatGPT}},
    journal         = {Machine Learning with Applications},
    volume          = {19},
    pages           = {no. 100622},
    year            = {2025},
    issn            = {2666-8270},
    doi             = {10.1016/j.mlwa.2025.100622},
    author          = {Yi Qi and Xingyu Zhao and 
                       Siddartha Khastgir and Xiaowei Huang},
}

@article{chen_trusta_2025,
    title           = {Trusta: Reasoning about assurance cases with formal methods and large language models},
    author          = {Chen, Zezhong and Deng, Yuxin and Du, Wenjie},
    journal         = {Science of Computer Programming},
    volume          = {244},
    pages           = {103288},
    year            = {2025},
    issn            = {1872-7964},
    doi             = {10.1016/j.scico.2025.103288},
    publisher       = {Elsevier}
}

@inproceedings{lipizzi_text_2025,
	title           = {From {Text} to {Structure}: {Extracting} and {Validating} {Complex} {System} {Representations} {Using} {Large} {Language} {Models}},
	author          = {Lipizzi, Carlo},
    booktitle       = {DS 141: Proceedings of the 27th International DSM Conference (DSM 2025)},
    organization    = {Hoboken, NJ, USA: the Design Society},
    url             = {https://www.designsociety.org/publication/48684/from_text_to_structure_extracting_and_validating_complex_system_representations_using_large_language_models},
	year           = {2025},
	pages          = {145--153},
}

@inproceedings{koh_retrieving_2025,
	title          = {Retrieving {Asymmetrical} {Indirect} {Links} {Through} {Large} {Language} {Models}},
	booktitle      = {DS 141: Proceedings of the 27th International DSM Conference (DSM 2025)},
	organization   = {Hoboken, NJ, USA: the Design Society},
	author         = {Koh, Edwin},
	year           = {2025},
    url            = {https://www.designsociety.org/download-publication/48688/retrieving_asymmetrical_indirect_links_through_large_language_models},
	pages          = {5--8},
}

@inproceedings{krus_augmenting_2025,
  title            = {Augmenting Aerospace System Design Using Large Language Models},
  author           = {Krus, Petter},
  booktitle        = {Proceedings of the 12th Swedish Aersopace Technology Congress - FT2025},
  organization     = {Stockholm, Sweden: Swedish Society for Aeronautics and Astronautics (FTF)},
  doi              = {10.3384/ecp215.1197},
  year             = {2025}
}

@article{liu_lostinmiddle_2017,
    title           = {Lost in the middle: How language models use long contexts},
    author          = {Liu, Nelson F and Lin, Kevin and Hewitt, John and Paranjape, Ashwin and Bevilacqua, Michele and Petroni, Fabio and Liang, Percy},
    journal         = {Transactions of the association for computational linguistics},
    volume          = {12},
    doi             = {10.1162/tacl_a_00638},
    pages           = {157--173},
    year            = {2024}
}

@article{chen2023extendingcontextwindowlarge,
    title           = {Extending Context Window of Large Language Models via Positional Interpolation}, 
    author          = {Shouyuan Chen and Sherman Wong and Liangjian Chen and Yuandong Tian},
    year            = {2023},
    journal         = {ArXiv: 2306.15595},
    eprint          = {2306.15595},
    archivePrefix   = {arXiv},
    primaryClass    = {cs.CL},
    doi             = {10.48550/arXiv.2306.15595}, 
}

@article{workslostinthemiddle_gupte_2025,
    title           = {What Works for 'Lost-in-the-Middle' in LLMs? A Study on GM-Extract and Mitigations}, 
    author          = {Mihir Gupte and Eshan Dixit and Muhammad Tayyab and Arun Adiththan},
    year            = {2025},
    journal         = {ArXiv: 2511.13900},
    eprint          = {2511.13900},
    doi             = {10.48550/arXiv.2511.13900},
    archivePrefix   = {arXiv},
    primaryClass    = {cs.CL}
}

@article{zhang2024middlelanguagemodelsuse,
      title         = {Found in the Middle: How Language Models Use Long Contexts Better via Plug-and-Play Positional Encoding}, 
      author        = {Zhenyu Zhang and Runjin Chen and Shiwei Liu and Zhewei Yao and Olatunji Ruwase and Beidi Chen and Xiaoxia Wu and Zhangyang Wang},
      year          = {2024},
      journal       = {ArXiV: 2403.04797},
      eprint        = {2403.04797},
      archivePrefix = {arXiv},
      primaryClass  = {cs.CL},
      doi           = {10.48550/arXiv.2403.04797}, 
}

@article{naveed2025comprehensive,
    title           = {A comprehensive overview of large language models},
    author          = {Naveed, Humza and Khan, Asad Ullah and Qiu, Shi and Saqib, Muhammad and Anwar, Saeed and Usman, Muhammad and Akhtar, Naveed and Barnes, Nick and Mian, Ajmal},
    journal         = {ACM Transactions on Intelligent Systems and Technology},
    volume          = {16},
    number          = {5},
    doi             = {10.1145/3744746},
    pages           = {1--72},
    year            = {2025},
    publisher       = {ACM New York, NY}
}

@article{zhang2026positionalfailureslongcontextllms,
    title           = {Positional Failures in Long-Context LLMs: A Blind Spot in Reasoning Benchmarks}, 
    author          = {Chuyifei Zhang and Hongyu Cui and Xiaowen Huang and Jitao Sang},
    journal         = {ArXiv: 2605.23170},
    year            = {2026},
    eprint          = {2605.23170},
    archivePrefix   = {arXiv},
    PrimaryClass    = {cs.CL},
    doi             = {10.48550/arXiv.2605.23170}, 
}

@article{baker2024lostmiddleinbetweenenhancing,
      title         = {Lost in the Middle, and In-Between: Enhancing Language Models' Ability to Reason Over Long Contexts in Multi-Hop QA}, 
      author        = {George Arthur Baker and Ankush Raut and Sagi Shaier and Lawrence E Hunter and Katharina von der Wense},
      year          = {2024},
      journal       = {ArXiV: 2412.10079},
      eprint        = {2412.10079},
      doi           = {10.48550/arXiv.2412.10079},
      archivePrefix = {arXiv},
      primaryClass  = {cs.CL},
      url           = {https://arxiv.org/abs/2412.10079}, 
}

@article{li2025ordermattersrethinkingprompt,
      title         = {Order Matters: Rethinking Prompt Construction in In-Context Learning}, 
      author        = {Warren Li and Yiqian Wang and Zihan Wang and Jingbo Shang},
      year          = {2025},
      journal       = {ArXiv: 2511.09700},
      eprint        = {2511.09700},
      doi           = {10.48550/arXiv.2511.09700},
      archivePrefix = {arXiv},
      primaryClass  = {cs.CL}
}

@inproceedings{guoetal2024makesgoodorder,
    title = "What Makes a Good Order of Examples in In-Context Learning",
    author = "Guo, Qi  and
      Wang, Leiyu  and
      Wang, Yidong  and
      Ye, Wei  and
      Zhang, Shikun",
    editor = "Ku, Lun-Wei  and
      Martins, Andre  and
      Srikumar, Vivek",
    booktitle = "Findings of the Association for Computational Linguistics: ACL 2024",
    month = aug,
    year = "2024",
    address = "Bangkok, Thailand",
    publisher = "Association for Computational Linguistics",
    doi = "10.18653/v1/2024.findings-acl.884",
    pages = "14892--14904",
}

@inproceedings{crabbjonesGenAI_2024,
    author          = {Crabb, Erin and Jones, Matthew T.},
    booktitle       = {2024 19th Annual System of Systems Engineering Conference (SoSE)}, 
    title           = {Accelerating Model-Based Systems Engineering by Harnessing Generative {AI}}, 
    year            = {2024},
    pages           = {110-115},
    doi             = {10.1109/SOSE62659.2024.10620975}
}

@article{conf:geisslerLLMAgent:2024, 
    title           = {Concept-Guided {LLM} Agents for {Human-AI} Safety Codesign}, 
    volume          = {3}, 
    DOI             = {10.1609/aaaiss.v3i1.31188}, 
    number          = {1}, 
    journal         = {Proceedings of the AAAI Symposium Series}, 
    author          = {Geissler, Florian and Roscher, Karsten and Trapp, Mario}, 
    year            = {2024}, 
    month           = {May}, 
    pages           = {100-104}
}

@inproceedings{dehn2025generating,
    title           = {Generating SysML V2 Models from Natural Language Requirements Using Large Language Models},
    author          = {Dehn, Simon and Schn{\"u}rer, Simon and Jacobs, Georg and H{\"o}pfner, Gregor},
    booktitle       = {2025 IEEE International Symposium on Systems Engineering (ISSE)},
    pages           = {1--7},
    doi             = {10.1109/ISSE65546.2025.11369988},
    year            = {2025},
    organization    = {IEEE}
}

@article{krus_FluidPowerLLMs_2026,
    title           = {Using Large Language Models for Fluid Power System Design},
    author          = {Krus, Petter},
    journal         = {JFPS International Journal of Fluid Power System},
    volume          = {19},
    number          = {2},
    pages           = {74--79},
    doi             = {10.5739/jfpsij.19.74},
    year            = {2026},
    publisher       = {The Japan Fluid Power System Society}
}

@inproceedings{mariniKrus_PromptComposition_2024,
    author          = {Marini, Vinicius Kaster and Krus, Petter},
    booktitle       = {10th CEAS Aerospace Europe Conference, 28th AIDAA International Congress},
    title           = {{Synthesizing aircraft system specifications with document-managed Large Language Model outputs
from one-shot system inquiry prompt chain.}},
    doi             = {10.21741/9781644904251-99},
    year            = {2025}
}

@article{qwen3,
    title        = {Qwen3 Technical Report},
    author       = {Yang, An and Li, Anfeng and Yang, Baosong and Zhang, Beichen and Hui, Binyuan and Zheng, Bo and Yu, Bowen and Gao, Chang and Huang, Chengen and others},
    journal      = {ArXiv:2505.09388},
    year         = {2025},
    doi          = {10.48550/arXiv.2505.09388}
}

@article{openai2025gptoss,
  title        = {{gpt-oss-120b} \& {gpt-oss-20b} Model Card},
  author       = {{OpenAI}},
  year         = {2025},
  journal      = {ArXiv: 2508.10925},
  eprint       = {2508.10925},
  archivePrefix = {arXiv},
  primaryClass = {cs.CL},
  doi          = {10.48550/arXiv.2508.10925},
}

@article{nemotron3super,
  title        = {Nemotron 3 Super: Open, Efficient Mixture-of-Experts Hybrid
                  Mamba-Transformer Model for Agentic Reasoning},
  author       = {{NVIDIA}},
  year         = {2026},
  journal      = {ArXiv: 2604.12374},
  eprint       = {2604.12374},
  archivePrefix = {arXiv},
  primaryClass = {cs.LG},
  doi          = {10.48550/arXiv.2604.12374},
  url          = {https://arxiv.org/abs/2604.12374}
}

@misc{ollama_kimi_k25_2026,
  title        = {Kimi K2.5 Model Card},
  author       = {{Ollama}},
  year         = {2026},
  howpublished = {Ollama model card},
  url          = {https://ollama.com/library/kimi-k2.5},
  note         = {Accessed: 2026-06-04}
}

@misc{sonnet46,
    title        = {Introducing {Claude Sonnet 4.6}},
    author       = {{Anthropic}},
    year         = {2026},
    month        = {February},
    howpublished = {Anthropic Product Announcement},
    url          = {https://www.anthropic.com/news/claude-sonnet-4-6}
}

@article{conf:vonheissenGenSysArch:2024,
    title           = {Toward Intelligent Generation of System Architectures},
    author          = {Von Heissen, Oliver and Hanke, Fabian and Mpidi Bita, Isaac and others},
    journal         = {Proceedings of NordDesign 2024},
    pages           = {504--513},
    url             = {https://www.designsociety.org/publication/47646/toward_intelligent_generation_of_system_architectures},
    organization    = {Reykjavik, Iceland: Design Society},
    year            = {2024}
}

\end{document}